\documentclass[english,aps,prl, amsmath,amssymb,twocolumn,letterpaper,showpacs,superscriptaddress]{revtex4-1}
\usepackage[T1]{fontenc}
\usepackage[latin9]{inputenc}
\usepackage{babel}
\usepackage{verbatim}
\usepackage{amsmath}
\usepackage{amssymb}
\usepackage{geometry}
\usepackage{epstopdf}
\usepackage{scrextend}

\usepackage{chngcntr}
\usepackage{caption}
\usepackage{ragged2e}
\DeclareCaptionJustification{justified}{\justifying}
\usepackage{floatrow}
\usepackage{graphicx}
\usepackage[label font={bf,normalsize}]{subfig}
\usepackage[unicode=true,pdfusetitle,
bookmarks=true,bookmarksnumbered=false,bookmarksopen=false,
breaklinks=false,pdfborder={0 0 1},backref=false,colorlinks=false]
{hyperref}
\usepackage{cleveref}

\usepackage{booktabs}

\AtBeginDocument{
\heavyrulewidth=.08em
\lightrulewidth=.05em
\cmidrulewidth=.03em
\belowrulesep=.65ex
\belowbottomsep=0pt
\aboverulesep=.4ex
\abovetopsep=0pt
\cmidrulesep=\doublerulesep
\cmidrulekern=.5em
\defaultaddspace=.5em
}
\makeatletter

\newcommand{\ket}[1]{\left| #1\right\rangle }
\newcommand{\bra}[1]{\left\langle #1 \right|}

\newcommand{\sgn}{\text{sgn}}

\newcommand{\abs}[1]{\left|#1\right|}
\newcommand{\down}{\downarrow}
\newcommand{\up}{\uparrow}
\newcommand{\pd}{\partial}

\newcommand{\sech}[1]{\text{sech}}
\newcommand{\cmnt}[1]{\ignorespaces}
\newcommand\numberthis{\addtocounter{equation}{1}\tag{\theequation}}
\usepackage[export]{adjustbox}
\begin{document}

\title{Universal quantum computing with parafermions assisted by a half fluxon}

\author{Arpit Dua}
\affiliation{Departments of Physics and Applied Physics, Yale University, New Haven, CT 06520, USA}
\affiliation{Yale Quantum Institute, Yale University, New Haven, CT 06520, USA}
\author{Boris Malomed}
\affiliation{Department of Physical Electronics, School of Electrical Engineering, Faculty of Engineering, Tel Aviv University, Tel Aviv, Israel}
\author{Meng Cheng}
\affiliation{Departments of Physics and Applied Physics, Yale University, New Haven, CT 06520, USA}
\author{Liang Jiang}
\affiliation{Yale Quantum Institute, Yale University, New Haven, CT 06520, USA}
\affiliation{Departments of Physics and Applied Physics, Yale University, New Haven, CT 06520, USA}

\date{\today}
\begin{abstract}
Braiding of anyons such as Majoranas or parafermions provides only Clifford gates which do not form a universal set of quantum gates. We propose a robust and resource-efficient scheme to perform a non-Clifford gate on a logical qudit encoded in parafermionic zero modes via the Aharonov-Casher effect. This gate can be implemented by moving a half flux quantum around the pair of parafermionic zero modes. The parafermion modes can be realized in a two-dimensional set-up using existing proposals and a half fluxon carrying half flux quantum can be created as a part of a half fluxon/anti-half fluxon pair in a spin-triplet Josephson junction with a dipole defect. With an appropriate bias current pulse, the half fluxon can be braided around the parafermions. Supplementing this gate with the braiding of parafermions provides the avenue for universal quantum computing with parafermions without magic state distillation.

\end{abstract}
\maketitle

\section{Introduction} 
Gottesman-Knill theorem~\cite{Gottesman_Knill} states that the quantum gates from the Clifford group can be efficiently simulated on a classical computer. Thus, in order to access the full computational power of quantum computers, one needs to go beyond the Clifford gates. In fact, one needs just a single non-Clifford gate~\cite{NC_proof} in order to densely generate the universal set of quantum gates. In the topological quantum computation (TQC) scheme~\cite{ROMP_TQC}, quantum information is stored in the non-local Hilbert space spanned by the so-called non-Abelian anyons that can emerge in topological phases of matter, and manipulated via the quantum gates generated by braiding of anyons. Examples of non-Abelian excitations include the Majorana zero modes (MZMs)~\cite{Moore91, Kitaev06a, Read00, Kitaev01,Alicea_review,FuKane}, their generalizations called parafermions (PFs)~\cite{FQH_setup, lindner2012, FTI_setup, barkeshli2012a, you2012} or even more exotic anyons called Fibonacci anyons~\cite{Read99}. Braiding of MZMs or PFs provides only the gates in the Clifford group. While braiding of Fibonacci anyons can provide the universal set of gates, their experimental realizations remain a major challenge~\cite{Read99, PF_review, Fibo_Mong_proposal, Fibo_platform_Alicea, Fibo_Vaezi_Z3}. On the other hand, experimental signatures for MZMs have been reported~\cite{Mourik2012, Experiment_Marcus, NichelePRL2017} and proposals made for braiding and error-correction~\cite{Alicea_rev_Majorana, Scalable_design_MF_MS}. Instead of physical braiding, one can perform measurement-based braiding~\cite{Heck, MBTQC,BondersonPRB2013, KnappPRX2016}, i.e., effective braiding via topological charge measurements with possible assistance from software~\cite{MOTQC_MFs,MOTQC_PFs} for improved efficiency. 

Topologically-protected parafermionic (PF) zero modes can be engineered as extrinsic defects in ``conventional'' Abelian topological phases~\cite{SET}, e.g. superconducting trenches in Fractional Quantum Hall (FQH)~\cite{FQH_setup} or Fractional Chern Insulator(FCI) systems, edge domain walls in Fractional Topological insulator~\cite{FTI_setup}, Fractional Topological Superconductor~\cite{Vaezi_2013}, lattice defects~\cite{you2012, teo2013, teo2013b}, or genons in bilayer FQH systems~\cite{barkeshli2012a, barkeshli2013genon}. A pair of PFs, for example in the FQH based set-up, has a composite topological charge that is a fraction $\frac{1}{N}$ of $2e$ electric charge where $N$ is an integer greater than 2. Hence, the associated qudit is immune to conventional quasiparticle poisoning which adds an integer multiple of $e$ to the system. This is unlike the systems for MZMs where $N$ is equal to 2 and hence suffer from quasiparticle poisoning~\cite{QPP_expt1,QPP_expt2, MF_code0,MF_code1,MF_code2}. Thus, if the fractional quasiparticle poisoning is suppressed, the PFs would hold an advantage over MZMs for the Clifford gates done via charge measurements. But still, like MZMs, gates based on braiding or topological charge measurements of PF modes lie in the Clifford group~\cite{QC_PFs}.

A key question for MZM/PF based TQC is implementation of a non-Clifford gate in order to have a universal gate set. For MZMs, there have been several proposals to implement the simplest qubit non-Clifford gate that belongs to the third level of Clifford hierarchy~\cite{qudit_NC}, the $\frac{\pi}{8}$ gate, via magic state distillations~\cite{Bravyi05}, tuning interactions between MZMs~\cite{MZM_Sarma_review,ROMP_TQC}, interferometry~\cite{ClarkePhaseGate} and universal geometric phase engineering~\cite{KarzigPRX2016}. For parafermions, the question is largely unexplored. It still remains an outstanding question to find a resource-efficient and robust protocol to implement a non-Clifford gate.

Qudit versions of the qubit non-Clifford gate like $\frac{\pi}{8}$ gate have been proposed~\cite{qudit_NC} and performance of magic state distillation protocols has been studied~\cite{NC_proof,qudit_NC_transversal}. In this work, we propose a robust method to implement a non-Clifford gate on a logical qudit encoded in parafermions via the Aharanov-Casher (AC) effect. Implementation of single-qubit unitary rotations for Majorana qubits using the AC effect has been discussed~\cite{Flux_read_out,TQB_parsa}. In~\cite{HF_chiral}, the current-phase relation for a Josephson junction made of spin-triplet superconductors has been calculated. We show that for such a Josephson junction, a half-fluxon(HF) is a solution for the order parameter phase difference across the junction. Braiding the HF around a pair of PF modes implements a non-Clifford gate on the associated qudit with dimension $N>2$. We investigate the HF solution for an annular spin-triplet Josephson junction, half-fluxon(HF)/anti-half-fluxon(AHF) pair creation in presence of a localized dipole current defect and calculate the bias current threshold for moving the HF. A pair configuration of localized HF (LHF) and free AHF (FAHF) is considered. A bias current pulse which ensures that the FAHF completes a single loop around the annular Josephson junction is constructed. Lastly, we discuss the robustness of the pair configuration and steps used in the gate implementation. 

\section{Non-Clifford gate and Implementation using Aharonov Casher effect}
In this section, we first define the particular non-Clifford gate of interest and then discuss its implementation using the AC effect. 
\subsection{Non-Clifford gate for qudits and parafermions}\label{NC_proof}
Pauli group~\cite{NC_proof} for a single qudit of dimension $N$ is defined as 
\[
P_{N}^{1}=\left\{X_{N}^{a}Z_{N}^{b}|a,b\in{0,1,...,N-1}\right\}
\numberthis
\]
and Pauli group for $m$ $N$-dimensional qudits is defined as $P_{N}^m=\left(P_{N}^{1}\right)^{\otimes m}$ where 
\[
X_{N}=\sum_{j=0}^{N-1}\ket{j\oplus1}\bra{j}, \hspace{2em} Z_{N} = \sum_{j=0}^{N-1}\omega^{j}\ket{j}\bra{j}.
\numberthis\label{eqn}
\]
Here, $\oplus$ is addition modulo $N$, $\omega = e^{\frac{i2\pi}{N}}$, and $j$ labels the computational basis. Clifford group for $m$ qudits is defined as $C_{N}^{m}=\left\{ U|\hspace{1em}UPU^{-1}\in P_{N}^{m} \hspace{1em} \forall P\in P_{N}^{m}\right\}$ as it preserves the Pauli group under conjugation.
We define $U_{\text{HF,N}}$ in the $N$-dimensional computational basis as a particular choice of the square root of $Z_N$ 
\[
U_{\text{HF,N}}=\text{diag}(1,e^{i\frac{\pi}{N}},e^{i2\frac{\pi}{N}}....,e^{i(N-1)\frac{\pi}{N}}). 
\numberthis\label{UHF}
\] 
Here, $HF$ denotes the half-fluxon since we use a half-fluxon to implement this gate. Conjugation of $X_{N}=\left(\begin{array}{cc}\mathbf{0}_{1,N-1} & 1\\ \mathbf{1}_{N-1,N-1} &\mathbf{0}_{N-1,1}\end{array}\right)$~\footnote{$\mathbf{0}_{M,N}$ represents a $M\times N$ null matrix and $\mathbf{1}_{M,N}$ represents a $M\times N$ identity matrix} by $U_{\text{HF,N}}$ gives
\[
U_{\text{HF,N}} X_{N}(U_{\text{HF,N}})^{-1}=e^{\frac{i\pi}{N}}\left(\begin{array}{cc}
\mathbf{0}_{1,N-1} & -1\\
\mathbf{1}_{N-1,N-1} & \mathbf{0}_{N-1,1}
\end{array}\right),
\numberthis\label{eqn12}
\]
which doesn't lie in the single qudit Pauli group for all $N>2$ (We excluded $N=2$ because in that case, the RHS of \cref{eqn12} reduces to $\sigma_y\in P_{2}^1$). Therefore, $U_{\text{HF,N}}$ is a non-Clifford gate for $N>2$.

\subsection{Aharonov-Casher effect}\label{NC_proof}
The Aharonov-Bohm(AB) effect~\cite{aharonov_bohm} in which a charge moving in a field-free region in a path enclosing a magnetic flux picks up a geometric phase, has a `dual' effect called the Aharonov-Casher(AC) effect~\cite{aharonov_casher}. In the AC effect, a neutral particle with a magnetic moment as it encircles an infinite line of charge picks up a phase proportional to the linear charge density. 

In a type-2 superconductor, if a charge $q$ braids around a localized fluxon in the bulk, it gets an AB phase. Aharonov and Reznick~\cite{reznik_aharonov} asked if it is possible to braid the fluxon around the charge $q$ instead to get an AC effect? Indeed, braiding a fluxon around a charge $q$ leads to accumulation of a geometric phase on the quantum state of the charge and the fluxon. In order to demonstrate the non-locality of the AC effect~\cite{reznik_aharonov}, the fluxon was considered to be in a force-free region i.e. a superconductor in which the electric field due to the charge is screened.  

Starting from a quantum state of a charge $q$ and flux tube with flux $\Phi$, $\ket{q}\otimes \ket{\Phi}$, braiding the flux tube with flux $\Phi$ around the charge $q$ gives a phase $e^{i\frac{\Phi q}{\hbar}}$ on the state. For a quantum state that is a charge superposition, each charge state gets a different AC phase leading to the implementation of a diagonal gate that is not proportional to Identity. Consider an example with $\mathbb{Z}_3$ parafermions. An arbitrary initial charge state of a pair of these parafermions can be written as 
\begin{equation}
\ket{\psi_{i}} =a_0\ket{0}+a_1\ket{\frac{2e}{3}}+a_2\ket{\frac{4e}{3}}
\label{charge_state}
\end{equation}
where $\ket{q}$ is a fractional charge state. The state after braiding a half quantum flux $\Phi$ around the parafermion pair due to gain of AC phase, $\phi_{AC}=\frac{q\Phi}{\hbar}$ on the fractional charge states $\{\ket{q}\}$ is 
\begin{equation}
\ket{\psi_{f}}= a_0\ket{0}+e^{i\frac{2e\Phi}{3\hbar}}a_1\ket{\frac{2e}{3}}+e^{i\frac{4e\Phi}{3\hbar}}a_2\ket{\frac{4e}{3}}
\label{gate}
\end{equation}
Choosing $\Phi$ to be a half flux quantum $\frac{h}{4e}$ implements the Non-Clifford gate $U_{HF,3}$ given in \cref{UHF}, on $\ket{\psi_i}$ .

\begin{figure}
\centering
\sidesubfloat[]{\includegraphics[width = 2.5in]{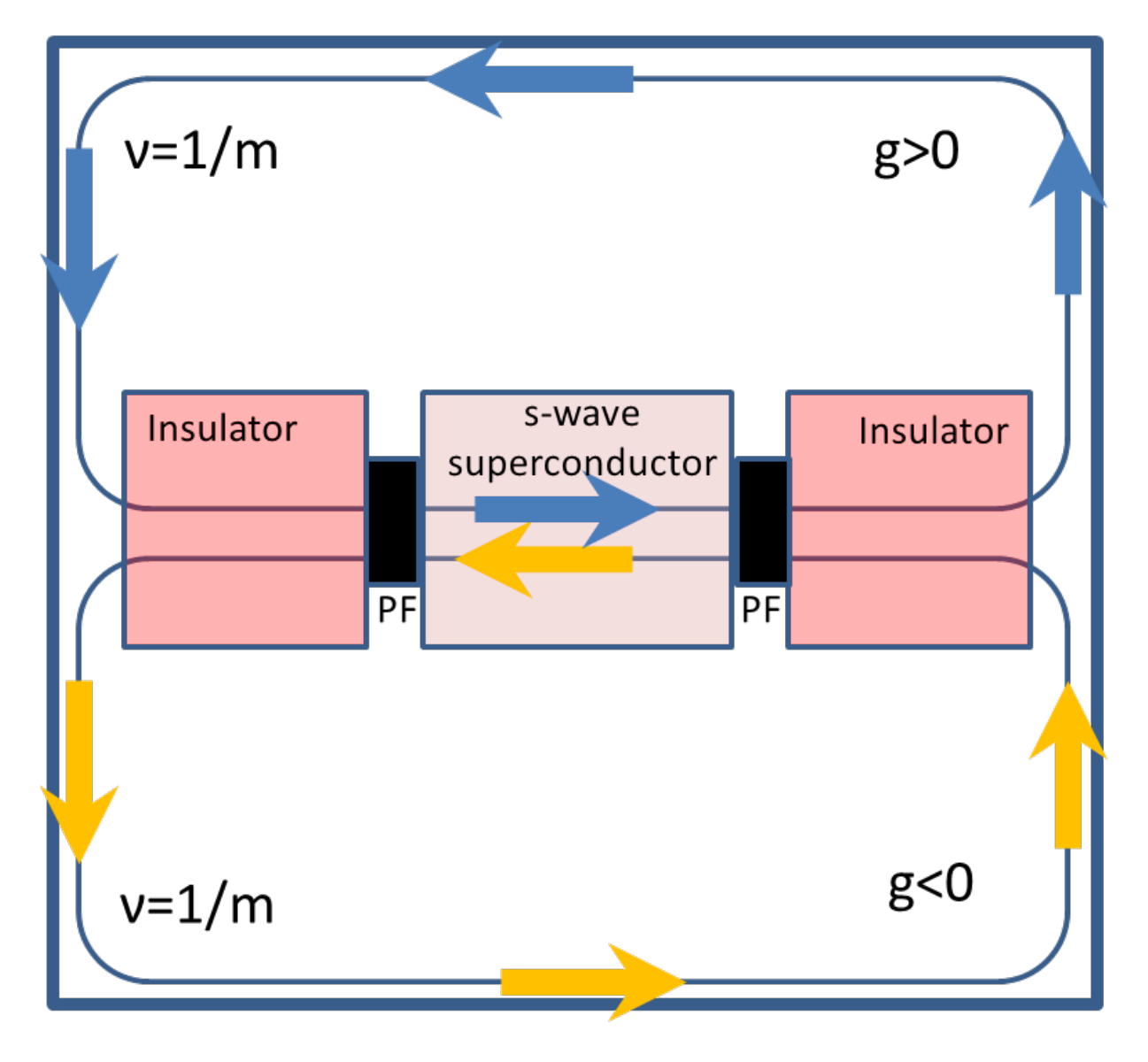}}\\ 
\sidesubfloat[]{\includegraphics[width = 2.5in]{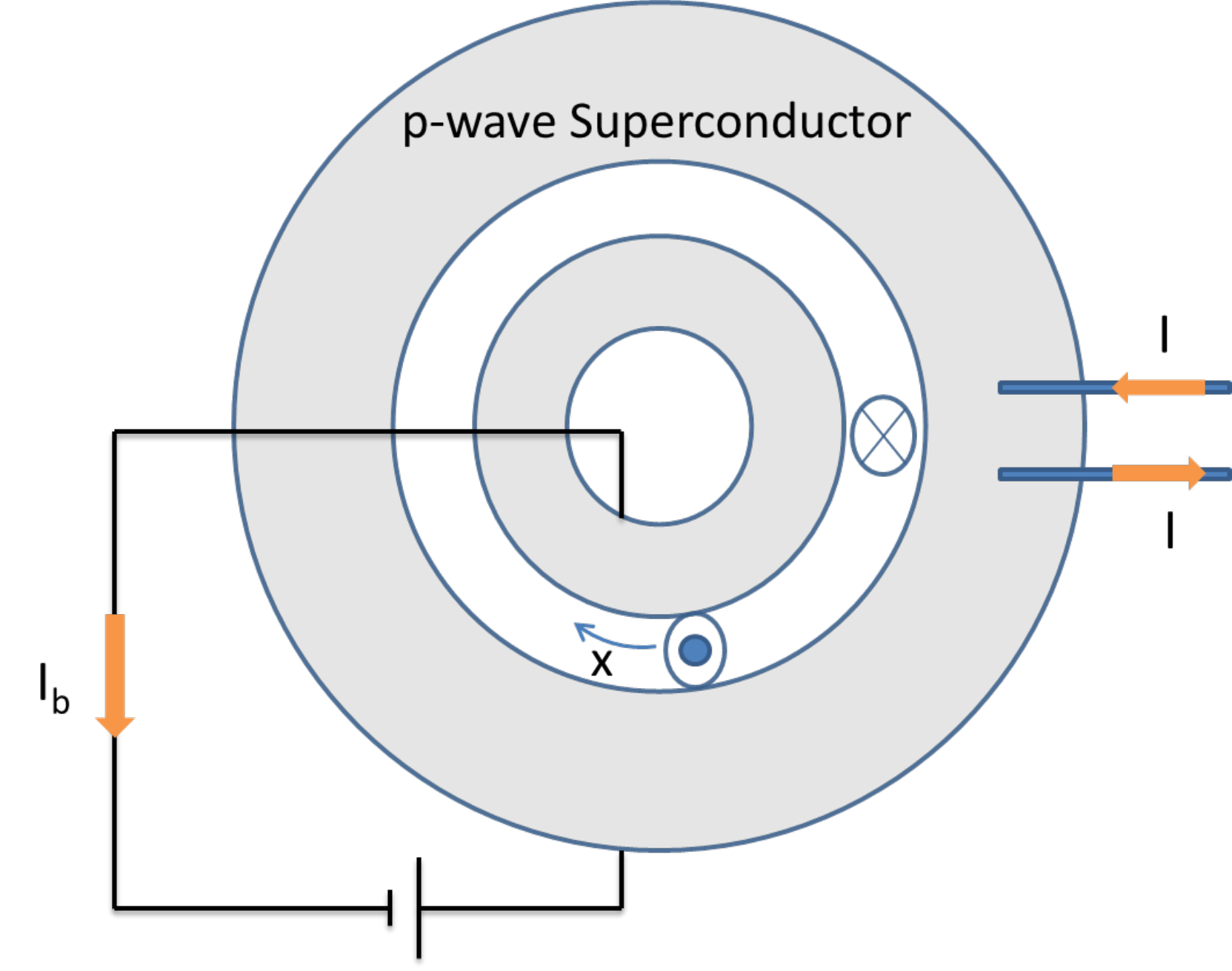}}\\ 
\sidesubfloat[]{\includegraphics[width = 2.5in]{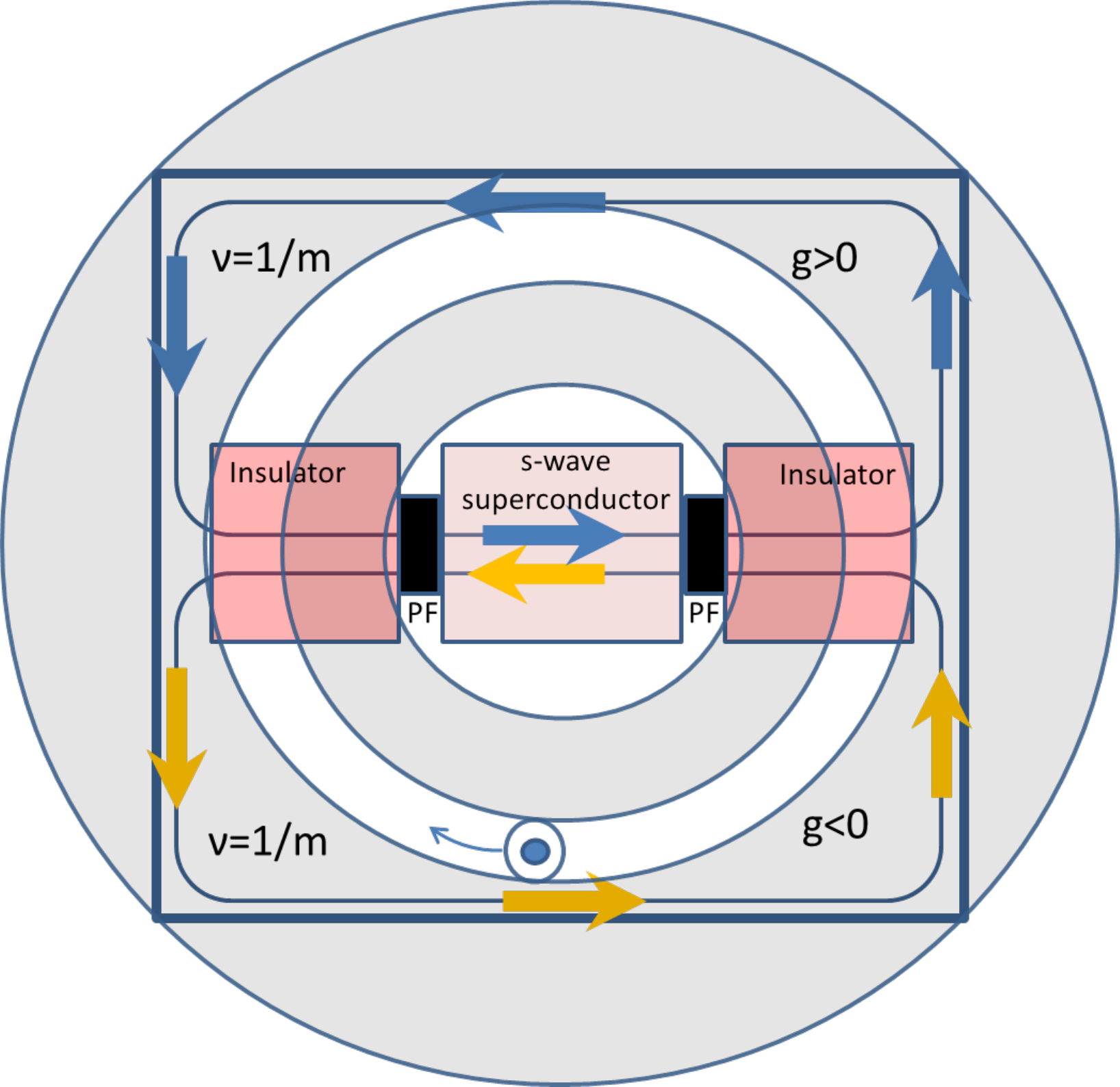}}
\caption{(a) Fractional Chern Insulator based setup for parafermionic (PF) defects shown by black patches at superconductor-insulator interface. (b) Spin-triplet Josephson junction with a localized dipole current that facilitates LHF-FAHF pair creation. The shaded ring-shaped regions present the superconductors while the region between them is the insulator. The bias current is shown as $I_b$, dipole current as $I$ and the distance between the injection and collection leads is marked as D. (c) Josephson junction setup combined with the FCI-based parafermionic setup~\cite{FQH_setup} such that the free AHF's flux can go around the pair of parafermionic defects to implement $\sqrt{Z_N}$ gate. Josephson junction has been placed on top of the setup with parafermionic zero modes. Here, the current injection leads as shown in (b) are not shown for simplicity.}
\label{PF_HF_setup}
\vspace{-0.7ex} 
\end{figure}

\section{Parafermionic set-up}
\begin{figure}
\centering
\sidesubfloat[]{\includegraphics[width = 2.5in]{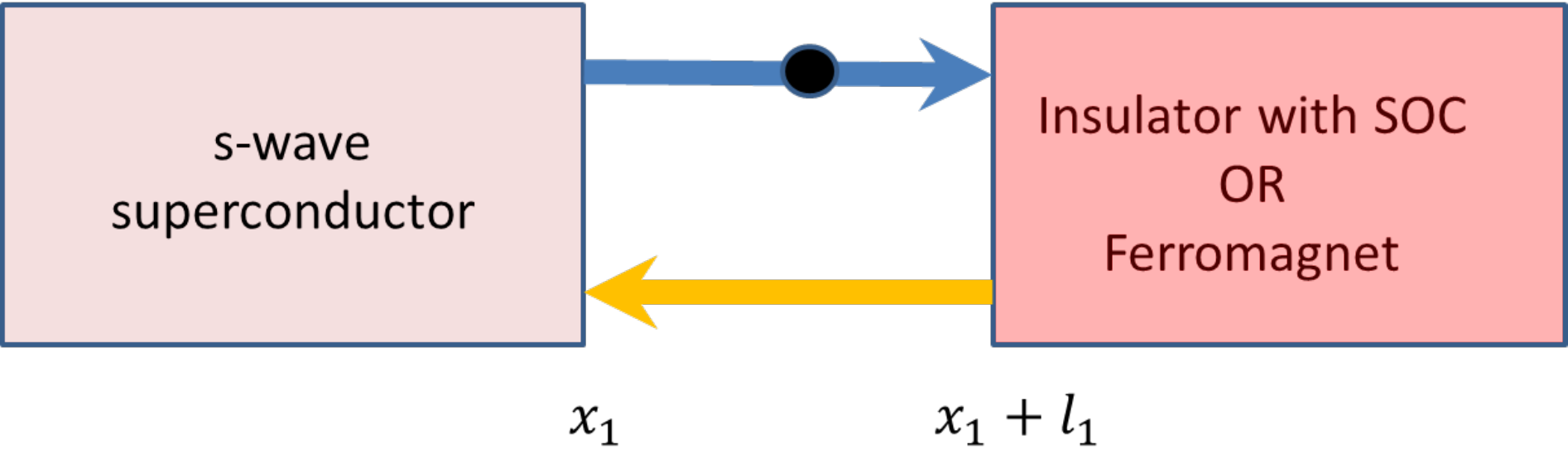}}\\ 
\sidesubfloat[]{\includegraphics[width = 2.5in]{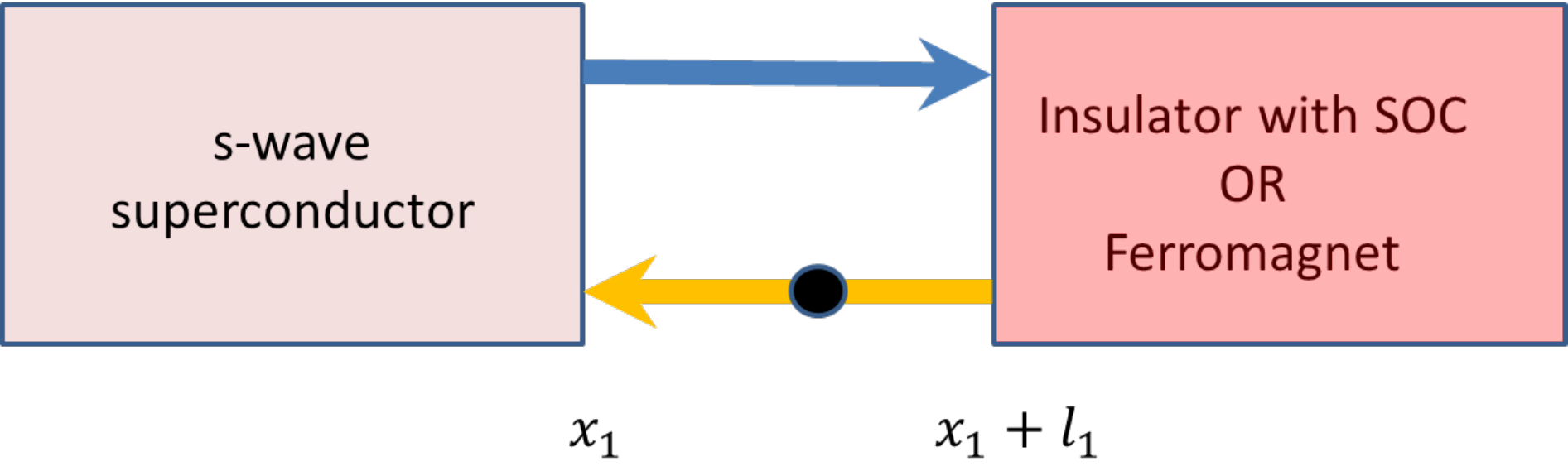}}\\ 
\sidesubfloat[]{\includegraphics[width = 2.5in]{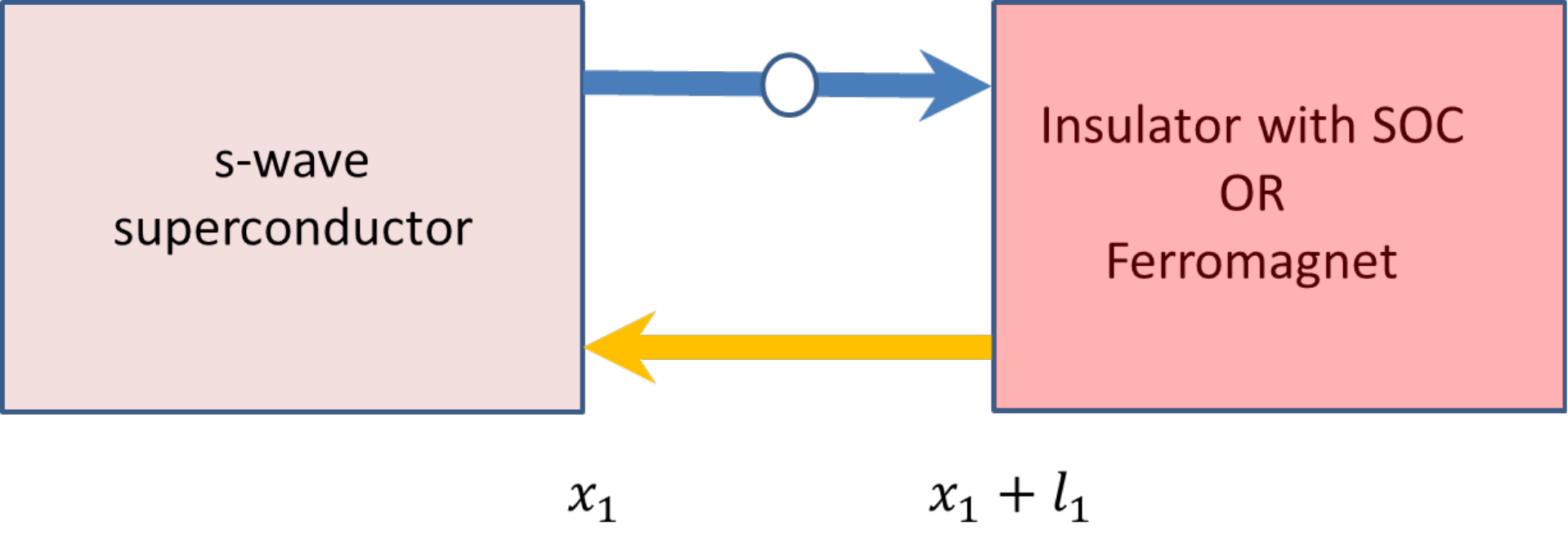}}\\ 
\sidesubfloat[]{\includegraphics[width = 2.5in]{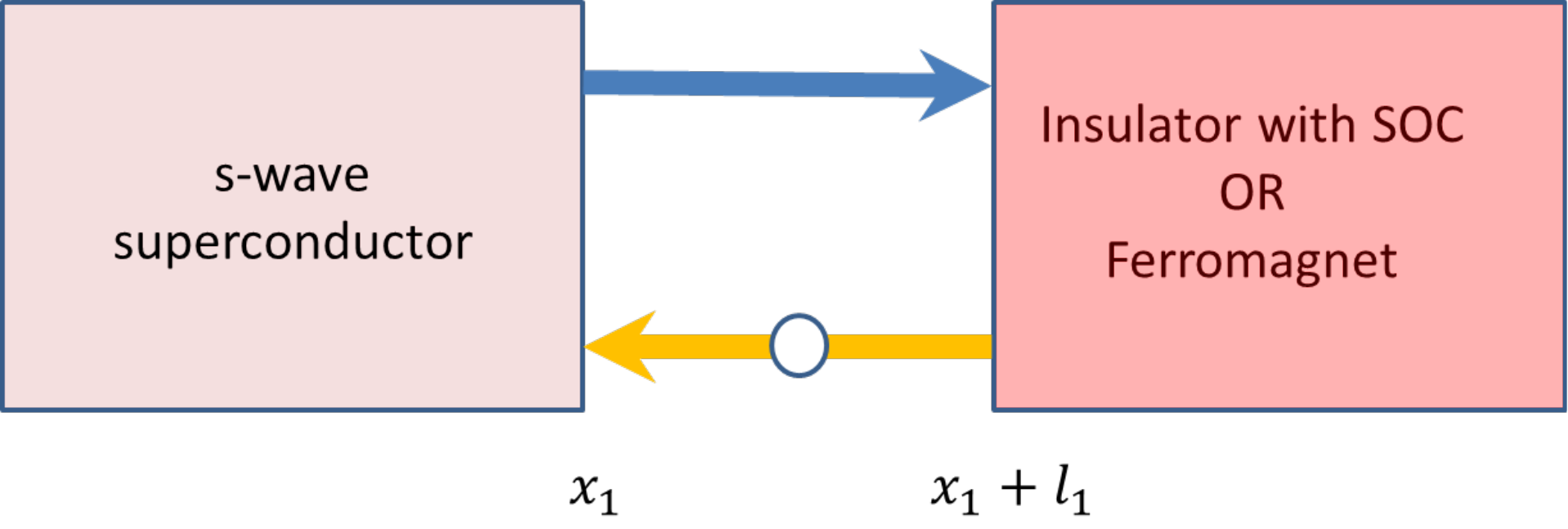}}\\ 
\caption{Processes that contribute to the parafermionic zero mode operator. Blue and orange arrows correspond to up and down spin while black and white circles correspond to quasiparticles and quasiholes respectively. $l_1$ is the length of the domain wall between the pairing and insulating region. (a) Right moving quasiparticle from pairing to insulating region. The insulator with spin-orbit coupling (SOC) or ferromagnet reflects the quasiparticle back with down spin in (b). (c) Andreev reflection from the pairing region sends back a quasihole with spin reversed back to up spin. (d) Quasihole gets reflected back from the insulating region to the pairing region with down spin.}
\label{process_zero_mode}
\vspace{-0.7ex} 
\end{figure}

Since our goal is to implement a non-Clifford gate using parafermions, we consider proposals for parafermionic defects from Ref.~\cite{FQH_setup,FTI_setup}. As discussed in these proposals, the first main ingredient is two adjacent Fractional Quantum Hall(FQH) wells or equivalently, two Fractional Chern Insulators as shown in Fig.~\ref{PF_HF_setup}, each with the same filling fraction $\nu=\frac{1}{k}$ where $k$ is an integer. The counter-propagating helical edge modes from the two quantum wells with opposite spins can be seen at the interface. For the purpose of this paper, we would use Fractional Chern Insulator layers (or a single Fractional Topological Insulator~\cite{FTI_setup}) in order to avoid the complications due to a strong magnetic field in the case of FQH wells. An s-wave superconductor is placed on top of a section of the interface to allow a pairing gap to open in that section. A spin-orbit coupled insulator or ferromagnet is used to create an insulating/magnetic gap in the neighboring sections. The domain walls between the pairing gap region and the insulating gap region form the parafermionic defects or zero modes. The tunneling of fractional charges between the domain walls is suppressed due to the pairing and insulating gaps. The zero mode operator $\zeta_1$ at the domain wall between the pairing region that ends at $x_1$ and insulating region that starts at $x_1+l_1$, is expressed as follows~\cite{FQH_setup, PF_review},

\begin{align*}
& \zeta_1\\
&= \int_{x_1}^{x_1+l_1}dx(\psi^\dagger_R(x)+\psi^\dagger_R(x_1+l_1)\psi_L(x_1+l_1)\psi^\dagger_L(x)+\\&  \psi^\dagger_R(x_1)\psi^\dagger_L(x_1)\psi^\dagger_R(x_1+l_1)\psi_L(x_1+l_1)\psi_R(x)+\psi^\dagger_R(x_1)\psi^\dagger_L(x_1)\psi_L(x))\\
& = \int_{x_1}^{x_1+l_1}dx(e^{i\phi_R(x)}+e^{i\left(\phi_R(x_1+l_1)-\phi_L(x_1+l_1)\right)} e^{i\phi_L(x)}+\\
& e^{i\left(\phi_R(x_1)+\phi_L(x_1)\right)}e^{i\left(\phi_R(x_1+l_1)-\phi_L(x_1+l_1)\right)}e^{-i\phi_R(x)}+\\ 
& e^{i\left(\phi_R(x_1)+\phi_L(x_1)\right)}e^{-i\phi_L(x)})
\numberthis\label{zeromode}
\end{align*}
where $\psi^\dagger_R(x)$ and $\psi^\dagger_L(x)$ are the creation operators for the right and left moving $e/m$ charges respectively at position $x$ on the edge of the FCI and in a bosonized framework, expressed as $\psi_{R/L}^\dagger(x)\sim e^{i\phi_{R/L}(x)}$ in terms of the fields $\phi_R(x)$ and $\phi_L(x)$. These fields obey the commutation relations $[\phi_{R/L}(x),\phi_{R/L}(x^\prime)]=\pm i\frac{\pi}{m}\sgn(x-x^\prime)$ and $[\phi_{L}(x),\phi_{R}(x^\prime)]=i\frac{\pi}{m}$. $\phi_{R/L}(x_1)$ and $\phi_{R/L}(x_1+l_1)$ are pinned due to the pairing and insulating gap terms respectively. The zero mode operator arises as a superposition~\cite{Alicea_review} of different processes, described by each of the terms in \cref{zeromode} and shown in Fig.~\ref{process_zero_mode}. The first term in \cref{zeromode},  $\psi^\dagger_R(x)$ corresponds to the right moving quasiparticle with up spin. The second term describes the reflection from the insulating region on which the quasiparticle reflects back to the pairing region with inverted spin. The third term describes the Andreev reflection from the pairing region due to which the left moving down spin quasiparticle gets converted into a right moving quasihole with up spin, whose creation operator is $\psi_R(x)$. The fourth term describes the quasihole reflected from the insulating region with inverted spin. This quasihole gets Andreev-reflected from the pairing region and converts to the right moving quasiparticle. The implementation of braiding of parafermions at the domain walls of a chain of superconducting and ferromagnetic islands is explained in Ref.~\cite{lindner2012} and equivalently in Ref.~\cite{FQH_setup}. 

\section{Spin-triplet superconductors and spin-triplet Josephson junction}
\label{HF_solution}
In this section, we first discuss the background on the spin-triplet superconductors and then the long spin-triplet Josephson junction. We show that for the long spin-triplet Josephson junction, there exists a half-fluxon (HF) solution. 

\subsection{Spin-triplet superconductors}
The spin-triplet superconductors are described by an order parameter matrix in spin and momentum space~\cite{HQV_stability2,helium3_Volhardt,SRO_review,evidence_spin_triplet} that can be expressed as
\[
\triangle\left(\mathbf{k}\right)=\left(\begin{array}{cc}
\triangle_{\up\up}\left(\mathbf{k}\right) & \triangle_{\up\down}\left(\mathbf{k}\right)\\
\triangle_{\down\up}\left(\mathbf{k}\right) & \triangle_{\down\down}\left(\mathbf{k}\right)
\end{array}\right)
\numberthis
\]
where $\bf{k}$ is the momentum and $\up$ or $\down$ indicates the z-component of spin. In general, we have 
\begin{align*}
    \triangle_{\sigma \sigma^{\prime}}\left(\mathbf{k}\right)=-\sum_{\mathbf{k^{\prime}}}V_{\mathbf{k,k^{\prime}}}\bra{G}f_{\mathbf{k}^{\prime}\sigma}^{\dagger}f_{-\mathbf{k}^{\prime}\sigma^{\prime}}^{\dagger}\ket{G}
    \numberthis
\end{align*}
where $\ket{G}$ is the ground state and the expression comes from applying mean field theory to the quartic fermionic interaction term $\sum_{\boldsymbol{k},\boldsymbol{k}^{\prime}}V_{\boldsymbol{k},\boldsymbol{k}^{\prime}}f_{\boldsymbol{k}\sigma}^{\dagger}f_{-\mathbf{k}\sigma^{\prime}}^{\dagger}f_{\boldsymbol{k}^{\prime}\sigma}f_{-\mathbf{k}^{\prime}\sigma^{\prime}}$ where $\sigma$ indicates up or down spin component, $f^\dagger_{{\bf{k}}\sigma}$'s are fermionic creation operators in momentum space and $V_{\bf{k},\bf{k^\prime}}$ is the Fourier coefficient of the interaction term. Thus, the superconducting order parameter $\triangle_{\sigma\sigma^{\prime}}\left(\mathbf{k}\right)$
is a wavefunction of a Cooper pair formed by two quasiparticles whose
momenta and spins are $\left(\mathbf{k},\sigma\right)$ and $\left(-\mathbf{k},\sigma^{\prime}\right)$. For a $p+ip$ superconductor, we can choose a spin coordinate system in which $\triangle_{\up\down}(\bf{k})=\triangle_{\down\up}(\bf{k})=0$ for all $\bf{k}$. For more details, look at appendix \ref{spin_triplet_bkg}. In this new coordinate system, we can write down the Hamiltonian of the spin-triplet $p_x+ip_y$ superconductor as 
\begin{align*}
    H&=\sum_{\boldsymbol{k}\sigma}\xi_{\boldsymbol{k}\sigma}f_{\boldsymbol{k}\sigma}^{\dagger}f_{\boldsymbol{k}\sigma}+\\
    &\frac{1}{2}\Big(\tilde{\triangle}_{\sigma\sigma}(\boldsymbol{k})^{\star}f_{-\boldsymbol{k}\sigma}f_{\boldsymbol{k}\sigma}+\tilde{\triangle}_{\sigma\sigma}(\boldsymbol{k})f_{\boldsymbol{k}\sigma}^{\dagger}f_{-\boldsymbol{k}\sigma}^{\dagger}\Big)
\numberthis\label{H_pwave}
\end{align*}
where $\xi_{\boldsymbol{k}\sigma}$ is the single-particle kinetic energy. $\tilde{\triangle}_{\sigma\sigma}(\boldsymbol{k})$ are the components of the order parameter matrix for the new choice of spin-quantization axis and given by  $\triangle_{\up\up}^{\star}\left(\boldsymbol{k}\right)=\frac{\triangle_{0}}{\sqrt{2}}e^{-i\Theta^\up_p}\left(k_{x}-ik_{y}\right)$
and $\triangle_{\down\down}^{\star}\left(\boldsymbol{k}\right)=\frac{\triangle_{0}}{\sqrt{2}}e^{-i\Theta^\down_p}\left(k_{x}-ik_{y}\right)$. Here, $\triangle_0$ is a constant and $\Theta^{\up(\down)}_p$ are the order parameter phases corresponding to the $\up(\down)$ spin component. Ground state of the above Hamiltonian \cref{H_pwave} can be written as
\[
\ket{G}=\prod_{\boldsymbol{k}\up}\left(u_{\boldsymbol{k}\up}+v_{\boldsymbol{k}\up}f_{\boldsymbol{k}\up}^{\dagger}f_{-\boldsymbol{k}\up}^{\dagger}\right)\prod_{\boldsymbol{k}^{\prime}\down}\left(u_{\boldsymbol{k}^{\prime}\down}+v_{\boldsymbol{k}^{\prime}\down}f_{\boldsymbol{k}^{\prime}\down}^{\dagger}f_{-\boldsymbol{k}^{\prime}\down}^{\dagger}\right)\ket 0 .
\numberthis\label{gs_pwave1}
\]
Here, $\frac{v_{\boldsymbol{k}\sigma}}{u_{\boldsymbol{k}\sigma}}=-\frac{\left(E_{\boldsymbol{k}\sigma}-\xi_{\boldsymbol{k}\sigma}\right)}{\triangle_{\sigma\sigma}^{\star}\left(\boldsymbol{k}\right)}$
where $E_{\boldsymbol{k}\sigma}=\sqrt{\xi_{\boldsymbol{k}\sigma}+\abs{\triangle_{\sigma\sigma}(\boldsymbol{k})}^2}$. 

\subsection{Spin-triplet Josephson junction}
The Hamiltonian for a conventional long Josephson junction~\cite{collective_coordinate,TopSC_JJ} can be generalized to the spin-triplet case as 
\begin{align*}
H= &\sum_{\sigma=\up,\down}\int dx\,\Big( \frac{1}{2}c_{n_{\sigma}}n_{\sigma}^{2}+ \frac{1}{2}c_{\sigma\sigma}\left(\pd_{x}\Theta^{\sigma}\right)^{2}\\
&+ \frac{1}{2}c_{\up\down}\left(\pd_{x}\Theta^{\up}\right)\left(\pd_{x}\Theta^{\down}\right)+J_{\sigma}\left(1-\cos\Theta^{\sigma}\right)-\frac{I_b}{2}\Theta^\sigma\Big),\\
\numberthis\label{eqn_model1}
\end{align*}
where $x$ is the coordinate along the length of the junction, $n_\sigma$ is the number charge density for spin $\sigma$ and $I_b$ is the bias current. $c_{\up\up}$ and $c_{\down\down}$ are the coefficients of the magnetic terms~\cite{collective_coordinate} and $c_{\up\down}$ is the coefficient of an allowed coupling term between the variation of $\Theta^\up$ and $\Theta^\down$. Last two terms are Josephson energy contributions from up and down spin sectors. $J_\sigma$ set the characteristic Josephson energy scales for the $\sigma$ spin component and $c_{n_\sigma}$ is the coefficient of the capacitive term. $\Theta^\sigma$ is differences of the order parameter phases across the junction ``seen'' by the $\sigma$ spin component. Under the assumptions $c_{\up\down}=0$, $c_{n_\up}=c_{n_\down}=c_n$, $c_{\up\up}=c_{\down\down}=c_\Theta$ and $J_\up=J_\down=J$, the equations of motion for this Hamiltonian can be written as
\begin{align*}
\frac{\pd_{{t}}^{2}{\Theta}^{\up(\down)}}{c_{n}}&=I_b-J\sin\Theta^{\up(\down)}+c_{\Theta}\pd_{x}^{2}\Theta^{\up(\down)}\\
\numberthis\label{eqnofmotion}
\end{align*}
For zero bias current, equations of motion are
\begin{align}
\pd_{\bar{t}}^{2}\Theta^{\up(\down)}+\sin\Theta^{\up(\down)}&=\pd_{\bar{x}}^{2}\Theta^{\up(\down)}
\end{align} 
where $\bar{x}=\frac{J}{c_\Theta}x$ and $\bar{t}=Jc_n t$. These equations have a traveling wave solution i.e. of the form  $\Theta^{\up(\down)}=\Theta^{\up(\down)}(\bar{x}-u\bar{t})$ given by 
\begin{align*}
    \Theta^\up&= 4 \arctan(e^{\pm\frac{\bar{x}-u\bar{t}}{\sqrt{1-u^2}}})\\
    \Theta^\down&=0,
    \numberthis
    \label{hf_soln}
\end{align*} 
where the parameter $u$ represents an arbitrary constant velocity of propagation. This is a half-fluxon solution since only $\Theta^\up$ jumps by $2\pi$. In appendix~\ref{half_fluxon_solution}, we show that this solution is associated with a magnetic flux of a half flux quantum. 

In section~\ref{defect}, we discuss how a localized dipole current defect can help facilitate tunnel creation of HF-AHF pair such that one of them, either HF or AHF, is localized at the dipole while the other one is free to move along the length of the junction. The localized dipole defect has an associated magnetic flux that is pinned and if the magnitude of this pinned flux attains the half-flux quantum, it will be energetically favorable to have this pinned flux compensated by a negative half-flux quantum. Depending on the relative orientation of dipole and HF/AHF, either HF and AHF can compensate the pinned flux. We choose a convention in which the compensating half-quantum flux is carried by HF. This would imply that an HF-AHF pair can be created in the junction such that HF's negative flux compensates the flux pinned at the defect while along the length of the junction at a distance $z$ from the defect, free AHF appears. Applying the bias current $I_b$ moves the free AHF along the length of the junction. We consider the defect potential in Sec.~\ref{HF_manipulation} in more detail.

\section{Braiding of half fluxon around the parafermions}
In order to braid an HF (or equivalently AHF) around the PF pair, the set-up supporting the PF defects should be combined with the device supporting the HF solution. In Fig.~\ref{PF_HF_setup}b, we show a schematic of the Josephson junction that supports the HF and it's superposition with the FCI based set-up having parafermionic defects~\cite{FQH_setup}. The superposition means putting the ring-shaped spin-triplet Josephson junction on top of the parafermion set-up such that the flux due to the HF in the junction, can penetrate the insulating region and circle around the parafermions. HF goes through the insulators and is sufficiently far from the superconductor so that the associated flux is not screened by it. Note that the superconductor in the Josephson junction is a p-wave spin triplet superconductor and the superconductor used for proximity effect in the parafermion set-up is an s-wave superconductor. It is the localized magnetic flux of the half-fluxon in the Josephson junction that braids around the pairing region leading to the following consequences.  

\subsection{HF winding around the s-wave superconductor}
As HF winds around the parafermion defects at the ends of a pairing region in the FCI setup, the associated flux passes around both the parafermionic modes, the pairing region of the FCI edges as well as the s-wave superconductor. When HF completes a loop around the s-wave superconductor, the order parameter phase of the superconductor, $\Phi_{\text{s}}$ changes by $\pi$ due to the Aharonov Casher effect. This follows from the fact that when a half flux quantum $\frac{h}{4e}$ is taken around a Cooper pair of charge $2e$, there is an AC phase of $\frac{\frac{h}{4e}2e}{\hbar}=\pi$ accumulated on the superconducting ground state. The BCS ground state wavefunction is given by
\[
\ket{G_{\text{BCS}}} = \prod_{\bf{k}} (u_{\bf{k}} +e^{i\Phi_{\text{s}}}e^{i\phi_{\bf{k}}}v_{\bf{k}} f^{\dagger}_{\bf{k}\uparrow}f^{\dagger}_{\bf{-k}\downarrow})\ket{0},
\numberthis\label{eqn}
\]
where $\Phi_{\text{s}}$ is the order parameter phase, $\ket{0}$ is the vacuum state and $\frac{u_{\bf{k}}}{v_{\bf{k}}}$ is determined by the BCS Hamiltonian. Due to the AC phase of $\pi$ on the Cooper pair part under winding by HF, the ground state becomes
\[
\ket{G_{\text{BCS}}^\prime} = \prod_{\bf{k}} (u_k +e^{i\Phi_{\text{s}}+\pi}e^{i\phi_k}v_k f^{\dagger}_{\bf{k}\uparrow}f^{\dagger}_{\bf{-k}\downarrow})\ket{0},
\numberthis\label{eqn}
\]
where the order parameter phase is now $\Phi_s+\pi$.

\subsection{Non-Clifford gate via braiding of half fluxon around parafermion pair}\label{NonCliffordAC}
Moving a half quantum flux around a pair of $\mathbf{Z_{N}}$ parafermions (here, $\mathbf{Z_{N}}$ refers to the symmetry group associated with the parafermion pair whose charge can take values in $\mathbf{Z_{N}}$) effectively implements the $U_{\text{HF,N}}$ gate of \cref{UHF}. We use the example of $\mathbf{Z}_3$ parafermions discussed before but keeping the FCI based set-up in mind. An arbitrary initial charge state superposition of the pairing region supporting $\mathbf{Z_{3}}$ parafermions can be written as 
\begin{equation}
\ket{\psi_{i}} =\alpha_0\ket{0}_{\Phi_s}+\alpha_1\ket{\frac{2e}{3}}_{\Phi_s}+\alpha_2\ket{\frac{4e}{3}}_{\Phi_s},
\end{equation}
where the fractional charge state $\ket{q}_{\Phi_s}$ is the state of the FCI edges in the pairing region, and can be expressed as $\ket{q}_{\Phi_s}=\sum_{n\in \mathbb{Z}}e^{i{\Phi_s} n} \ket{q+2en}$ which shows the dependence on the s-wave superconductor's order parameter phase $\Phi_s$. 

Besides the AC phase gain as shown in \cref{gate}, under HF winding, the order parameter phase $\Phi_s$ defined above also changes by $\pi$ due to the AC effect as shown in the next subsection. Hence, the state after braiding can be written as 
\begin{align*} 
& \ket{\psi_{f}}\\
=  & \alpha_{0}\ket 0_{\Phi_{s}+\pi}+e^{i\frac{\pi}{3}}\alpha_{1}\ket{\frac{2e}{3}}_{\Phi_{s}+\pi}+e^{i\frac{2\pi}{3}}\alpha_{2}\ket{\frac{4e}{3}}_{\Phi_{s}+\pi},
\numberthis
\end{align*}
where the action of the unitary $U_{\text{HF,N}}$ along with phase shift of $\Phi_s$ by $\pi$ takes the system to a different ground state manifold. 

\subsection{Non-Clifford gate that restores the order parameter}\label{OP_phase_NC}
\begin{figure}
\centering
\includegraphics[scale=0.4]{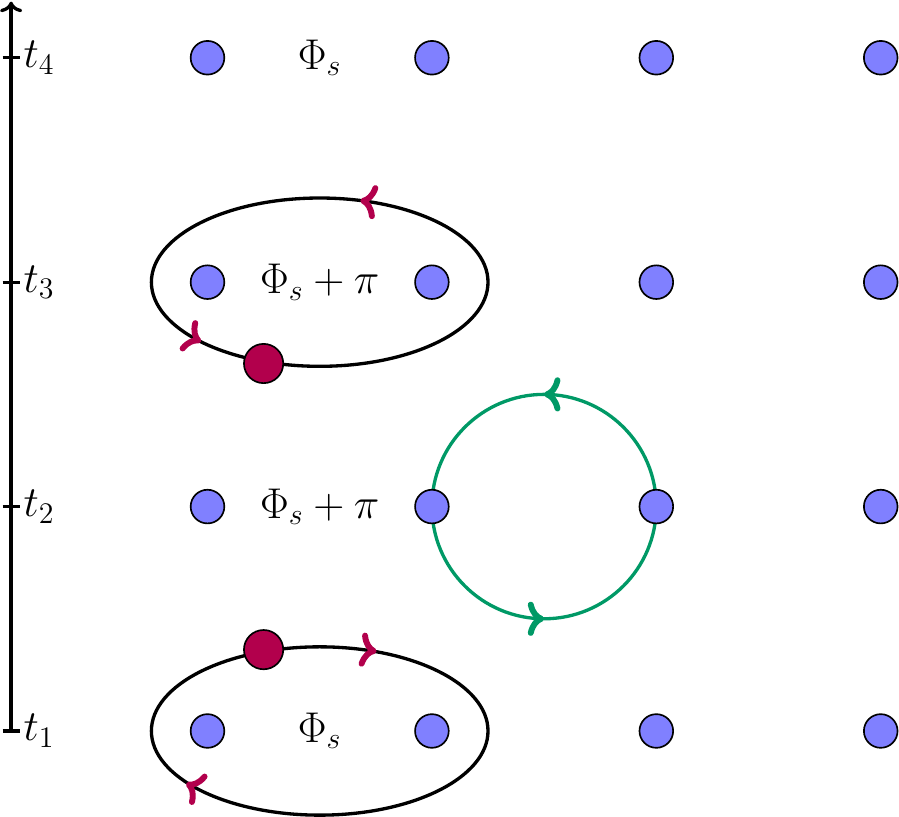}
\vspace{3mm}
\caption{Gate sequence composed of half-fluxon braidings in opposite directions (at time steps $t_1$ and $t_3$) with a parafermion braiding operation (at time step $t_2$ and shown in green), implements a non-Clifford operation. Red circle denotes the half fluxon and blue circles denote the parafermions. The first half-fluxon braiding induces a change of s-wave superconductor's order parameter phase from $\Phi_s$ to $\Phi_s+\pi$. The overall operation restores the order parameter phase back to $\Phi_s$ and hence, acts on the original ground state manifold.}
\label{HF_schematic}
\vspace{-5mm}
\end{figure}

We assume that the gate $U_{\text{HF,N}}$ described above uses braiding of HF in the clockwise direction. Braiding HF in the anticlockwise direction implements the unitary $(U_{\text{HF,N}})^{-1}$, but also shifts the order parameter phase of the s-wave superconductor $\Phi_s$ by $-\pi$. Using this, we find that an overall non-Clifford operation that restores the order parameter phase on the s-wave superconductor, can be achieved by inserting between the two HF braidings in opposite directions, a particular braiding operation, $U_B$. The combined evolution preserves the ground state manifold and is given by
\[
(U_{\text{HF,N}})^{-1}U_{B}U_{\text{HF,N}},
\]
which is a non-Clifford operation (for $N>2$). The braiding operation $U_B$ works for the new ground state manifold just like the original ground manifold because the operator content of the zero mode operators, in terms of which the braiding operation is defined, remains the same if the fields in the pairing and insulating region are still perfectly pinned i.e. the quasiparticle tunneling between the parafermion modes is suppressed because of the gap in the insulating region. For a logical qudit composed of 4 parafermions with fixed parity and the qudit state defined by the parity of first two parafermions $\gamma_1$ and $\gamma_2$, the operation $U_B$ can be chosen to be the braiding of parafermions $\gamma_2$ and $\gamma_3$. This is diagrammatically shown in Fig.~\ref{HF_schematic}. 

\section{Creation and manipulation of half-fluxon}
\label{HF_manipulation}
In this section, we discuss how to create an HF/AHF pair in a spin-triplet Josephson junction such that the HF is localized and the AHF is free to move around under the application of the bias current. We then find a bias current threshold below which the localized HF (LHF) remains localized. Subsequently, we design a bias current pulse such that the half-fluxon is free to move around. 

\subsection{Set-up design with defect potential to create HF/AHF pair}
\label{defect}
We use the set-up as described in~\cite{setup} but with an annular spin-triplet Josephson junction  instead of the conventional Josephson junction. In this set-up, in addition to the bias current $I_b$ considered earlier, we have an extra defect potential due to the injected localized dipole current $I$ as shown in Fig.~\ref{PF_HF_setup}a. The potential due to the current dipole or defect located near $x=0$ is given by 
\[
V(x,\phi) =\frac{I}{\lambda_J} (\delta(x-a)-\delta(x+a))\phi \numberthis \label{defect},
\]
where $\phi=\frac{\Theta^\up+\Theta^\down}{2}$ is the phase that couples to the magnetic vector potential as shown in appendix \ref{half_fluxon_solution}. In the limit of $a\rightarrow 0$, the potential given by \cref{defect} becomes $V(x,\phi)=-\varepsilon \delta^\prime(x)\phi$. Here, $a=\frac{D}{\lambda_J}$ where $D$ is the spacing between injection and collection leads and $\lambda_J$ is the Josephson penetration depth for the spin-triplet Josephson junction.  $\varepsilon=2\frac{I}{\lambda_J} a$ is the defect strength where $I$ is the injection current. Such an interaction was studied in~\cite{defect_AV} where the defect represented an Abrikosov vortex crossing the long Josephson junction. As mentioned in the previous section, the defect facilitates creation of an HF-AHF pair such that the HF compensates the magnetic flux of the pinned defect while the AHF is free to move along the length of the junction. We consider the creation of such an HF-AHF pair via under-barrier tunneling, starting from a vacuum configuration. In the absence of the defect due to localized current, the vacuum configuration is simply $\Theta^{\up(\down)}=0$. While in the presence of the defect, due to the boundary conditions for the phases $\Theta^\up$ and $\Theta^\down$ across the junction, shown in \cref{pert_bc_eqn}, one of the vacuum configurations mentioned in appendix~\ref{inhom_vac} can be achieved depending on the defect strength. Starting from the inhomogeneous vacuum configuration, we consider an instanton pair solution, that under imaginary time evolution i.e. under the energy barrier, ends up on the mass shell as a pair configuration of HF and AHF. We find a critical value of separation between HF and AHF, $z$ at which the pair configurations becomes on shell. Since applying a bias current moves the free HF around and increases the separation between HF and AHF, we expect that the bias current makes the pair-creation more favorable. We discuss the tunnel creation of HF-AHF pair and critical separation for on-shell condition in the appendix~\ref{pair_creation}. Now we discuss the bias current threshold and the bias current pulse such that the free AHF makes a single loop around the annular junction. 

\subsection{Bias current threshold}\label{threshold_section}
The boundary conditions for $\Theta^{\up(\down)}$ and $\Theta^{\up(\down)}$, found by integrating the equations of motion~\cref{eqn_defect} from appendix~\cref{inhom_vac}, are 
\begin{align*}
\Theta^{\up(\down)}(x=+0)-\Theta^{\up(\down)}(x=-0)=-\varepsilon\\
\pd_x \Theta^{\up(\down)}(x=+0)-\pd_x \Theta^{\up(\down)}(x=-0)=0.
\numberthis\label{pert_bc_eqn}
\end{align*} 
As mentioned before, the spin-triplet Josephson junction we consider has an annular shape as shown in Fig.~\ref{PF_HF_setup}b. For an annular Josephson junction, the Hamiltonian density ${\cal{H}}$ obeys the periodic boundary condition in the angular coordinate $x$ as 
\begin{align*} 
{\cal{H}}(\Theta^{\up(\down)}(x=-0))={\cal{H}}(\Theta^{\up(\down)}(x=+0)) \numberthis\label{eqn_ham}.
\end{align*} 
where $\Theta^{\up(\down)}(x=-0)$ is the value of $\Theta^{\up(\down)}$ to the left of defect's location $x=0$. Note that in the presence of the defect, the HF solution is modified to a localized HF solution shown in \cref{HFsoln} in appendix~\ref{inhom_vac}. This solution is periodic modulo $2\pi$ along the length of the annular junction and obeys the above boundary conditions. Denoting $\Theta^{\up(\down)}(x=-0)$ as $\Theta^{\up(\down)}_{-}$ and $\Theta^{\up(\down)}(x=-0)=\Theta^{\up(\down)}(x=+0)$ as $\Theta^{\up(\down)}_+$, \cref{eqn_ham} gives
\begin{align*}
&J(1-\cos\Theta^{\up}_-)+J(1-\cos\Theta^{\down}_-)-\frac{I_b}{2}(\Theta^{\up}_- + \Theta^{\down}_-) \\
=& J(1-\cos\Theta^{\down}_+)+J(1-\cos\Theta^{\down}_+)-\frac{I_b}{2}(\Theta^{\up}_+ + \Theta^{\down}_+).
\numberthis\label{eqn_thr}
\end{align*}
Using \cref{pert_bc_eqn} in \cref{eqn_thr} gives  
\begin{align*}
I_b & =\varepsilon^{-1}\sum_\sigma J\Big(\cos(\Theta^\sigma_- -\varepsilon)-\cos\Theta^\sigma_- \Big). \numberthis\label{eqn}
\end{align*}
The threshold value for the bias current, $I_b^\text{th}$ is found by varying $I_b$ w.r.t. both $\Theta^\up_-$ and $\Theta^\up_+$ and is given by 
\begin{align*}
I_b^\text{th}
=\pm 4\varepsilon^{-1}\sin\frac{\varepsilon}{2}.
\numberthis\label{eqn} 
\end{align*} 
as a function of the defect strength $\varepsilon$. 

\subsection{Bias current pulse for half fluxon loop}\label{bias_pulse}

A bias current pulse for braiding an FAHF around the junction can be designed using the collective coordinate picture~\cite{collective_coordinate} as follows. Multiplying the equation of motion \cref{eqnofmotion} for $\Theta^\up$ by $\Theta^\up_x$ and integrating, we get 
\[
\frac{1}{Jc_{n}}\int_{x}\Theta^\up_{x}\ddot{\Theta^\up}=I_b\int_{x}\Theta^\up_x-\int_{x}\Theta^\up_{x}\sin \Theta^\up-\frac{{c}_{\Theta}}{J}\int_{x}\Theta^\up_{x}\Theta^\up_{xx}. 
\numberthis\label{eqn}
\]
Plugging in the FAHF solution $\Theta^\up(x,t)= 4 \arctan(\frac{-x+z}{\beta})$ where $\beta=\sqrt{1-(\frac{dz}{dt})^{2}}$, performing the integration, we get
\[
\frac{1}{J c_{n}}\int_{x}\Theta^\up_{x}\ddot{\Theta^\up}=-2I_b\pi.
\numberthis\label{eqn}
\]
Taking the velocity of FAHF, $\frac{dz}{dt}$ to be a constant and identifying the mass of the HF as $M=\frac{1}{2\pi J c_{n}}\int dx\Theta_{x}^{2}=\frac{4}{\pi J\beta c_{n}}$, we get 
\[
M\ddot{z}=I_b,
\numberthis\label{eqnHF}
\]
which is an analog of Newton's equation of motion of AHF of mass $M$ with collective coordinate $z(t)$. Using this equation of motion, we can design a bias current pulse $I_b(t)$ such that the AHF goes around the junction once and comes to a stop. The force of attraction of AHF with the pinned HF is expected to decay exponentially with the HF-AHF separation and we do not consider it in this calculation. In principle it can be taken into account and the required pulse can be found numerically. The boundary conditions for the AHF that moves along the circular Josephson junction, are $z(0)=0$, $\dot z(0)=0$, $z(T)=2\pi$, $\dot z(T)=0$ where $T$ is the time for completing the loop and can be chosen. Denoting $I_b$ as a function of both $t$ and $T$, we need to check that the bias current at all times is below the threshold value,
\[
\abs{I_b(t)}<\abs{I_b^\text{th}}
\numberthis
\label{threshold_cond}
\]
so that the LHF remains localized at the location of the defect. Writing $I_b(t)=\dot Q(t)$ where $Q(t)$ is the bias charge and integrating \cref{eqnHF} over time twice, we get 
\[
\dot z(t)=\frac{1}{M}(Q(t)-Q(0)). \numberthis\label{eqn}
\]
Boundary condition $\dot z(0)=0$ is satisfied and using boundary condition $\dot z(T)=0$, we get $Q(T)=Q(0)$. Integrating once more, we get 
\[
z(t)=\frac{1}{M}(\int_{0}^{t}dt^{\prime}Q(t^\prime)-tQ(0)). \numberthis\label{eqn}
\]
Boundary condition $z(0)=0$ is satisfied and $z(T)=2\pi$ gives $\frac{\pi}{M}(\int_{0}^{T}dt^{\prime}Q(t^\prime)-TQ(0))=2\pi$. Choosing $Q(t)=a(1-\cos(\Omega t))+bt$ where $\Omega$ is arbitrary parameter that can be chosen. Note that $Q(0)=0$. On satisfying the boundary conditions, we find the coefficients $a$ and $b$ to get the required bias current pulse as
\begin{align*}
I_b(t,T)&=\frac{4\pi M \Omega }{T \Big(\cos (T \Omega )+1-2 \frac{\sin(T \Omega )}{T\Omega}\Big)}\sin(\Omega t)\\ 
   &+\frac{4 \pi  M (\cos (T \Omega )-1)}{T^2   \Big(\cos (T \Omega
   )+1-2\frac{ \sin (T \Omega )}{T \Omega}\Big)}
\numberthis\label{eqn:bias pulse} 
\end{align*}
where $\Omega$ and $T$ can be tuned to satisfy the condition~\cref{threshold_cond}. 

\section{Discussion}\label{Robustness}
We first state the assumptions made in our protocol of the implementation of non-Clifford gate. We assume that the HF braiding doesn't affect the state of the spin-triplet Josephson junction or the superconductors in the spin-triplet Josephson junction. This is also supported from the fact that spin-triplet superconductors support half quantum vortices~\cite{ivanov} in the bulk and braiding such a vortex around a region of the bulk shouldn't change the ground state. The detailed analysis of this is beyond the scope of this work. 

Secondly, we assume that different pairing regions in a set-up that supports a chain of parafermionic defects, can have different order parameter phases. This leads to a higher ground state energy but as long as the quasiparticle tunneling between the domain walls is suppressed due to the insulating gap and the spacing between them, the form of the parafermion operators at the domain walls retain the same dependence on the pinned fields in the neighboring pairing and insulation regions. Hence, the Josephson effect due to the difference of order parameter phases can lead to tunneling of only Cooper pairs and that would also be suppressed due to the spacing between domain walls. Third, in this work, we have ignored technical issues that may arise in combining the parafermion set-up with the spin-triplet Josephson junction such that one of them is on top of the other one. 

Lastly, as the HF is being braided, a fractional quasi-particle from the FCI layer could get trapped in it. But we assume that it won't be able to cross the interface as that process will be energetically suppressed. Even if a quasi-particle gets trapped and participates along with HF, in braiding around the overall abelian charge on the island, it will lead to an extra overall Clifford gate in addition to the non-Clifford gate from HF braiding. The overall gate will still be non-Clifford but it will be ambiguous up to a Clifford gate. Such ambiguity in the application of non-Clifford gate is characteristic of simple non-abelian systems for topological quantum computing and hence, quasi-particle trapping needs to be controlled for or kept track of in some manner. 

Now we discuss how well-controlled can the HF/AHF pair-creation process be made. In our implementation, we need an HF/AHF pair to be created via under-barrier tunneling, such that HF is localized and AHF is free to move (or vice-versa). Interaction of HF alone, with the defect, given by $-\int_x\varepsilon \delta^\prime(x)\phi_{\text{HF}}$, is equal but opposite to that of AHF with the defect. Hence under tunnel creation, one of them comes to the mass shell in the pinned state while the other tends to be free~\cite{pair_creation}. In appendix~\ref{pair_rates}, we calculate the pair creation rate for the LHF/FAHF pair and also for the fluxon pair of localized fluxon (LF) and free antifluxon (FAF). The pair creation is considered on top of an inhomogeneous quadrupole vacuum configuration which can be achieved by tuning the defect strength. We show that the pair creation rate of LF/FAF pair is exponentially suppressed compared to the LHF/FAHF pair creation rate. 

\section{Conclusions}
We proposed a non-Clifford gate for parafermions using Aharonov-Casher effect. Braiding a half-fluxon around a parafermion pair implements the $U_{\text{HF,N}}$ gate as mentioned in \cref{UHF} and which is non-Clifford for qudit dimension $N>2$. In a spin-triplet Josephson junction with a current dipole defect, half-fluxon can be created and moved around using a bias current. Combining such a junction with parafermionic defects can implement the non-Clifford gate robustly via half-fluxon braiding. This proposal can be combined with recent work on parafermion box \cite{PF_box} for a universal gate set where the Clifford gates are measurement-based. While we have focused on half-fluxons in spin-triplet superconductors to develop a proof-of-principle scheme, the key ingredient, namely the existence of a stable vortex with fractional flux, may appear in other types of systems, for example, unconventional superconductors intertwined with spatial order~\cite{Berg09, SarangPRL2013}. By braiding a quantized fractional flux of a quarter fluxon i.e. $\frac{h}{8e}$ around a pair of Majorana zero modes, one can also implement the non-Clifford gate $T=Z^\frac{1}{4}$ on the corresponding logical qubit with logical operator $Z$. Thus, extending anyon models with quantized fractional fluxons provides robust universality and a study of such extensions is left for future work. 

\section*{Acknowledgments}
\begin{acknowledgments}
AD thanks Stephan Plugge for enlightening discussions. AD and LJ also thank Yuval Oreg, Ady Stern, Jason Alicea, Leonid Glazman, Roman Lutchyn, Chetan Nayak and Maissam Barkeshli for discussions. AD thanks Jukka V\"ayrynen for comments. This work was supported by the ARL-CDQI, ARO (Grants No. W911NF-18-1-0020 and No. W911NF-18-1-0212), ARO MURI (W911NF-16-1-0349), NSF (EFMA-1640959), AFOSR MURI (FA9550-14-1-0052 and FA9550-15-1-0015), the Alfred P. Sloan Foundation (BR2013-049), and the Packard Foundation (2013-39273).
\end{acknowledgments}
\vspace{-2em}
\bibliographystyle{apsrev4-1}
\bibliography{refs1}

\appendix

\onecolumngrid
\setcounter{equation}{0}
\setcounter{figure}{0}
\setcounter{table}{0}

\section{Background on spin-triplet superconductors}
\label{spin_triplet_bkg}
We reiterate the facts covered in main text on spin-triplet superconductors with more explanation. As mentioned, the spin-triplet superconductors are described by an order parameter matrix in spin and momentum space~\cite{HQV_stability2,helium3_Volhardt} as follows-
\[
\triangle\left(\mathbf{k}\right)=\left(\begin{array}{cc}
\triangle_{\up\up}\left(\mathbf{k}\right) & \triangle_{\up\down}\left(\mathbf{k}\right)\\
\triangle_{\down\up}\left(\mathbf{k}\right) & \triangle_{\down\down}\left(\mathbf{k}\right)
\end{array}\right)
\numberthis
\]
where the arrows indicate spin quantum number of each electron in the pair, $s_z$. The order parameter matrix elements are given by
\begin{align*}
    \triangle_{\sigma \sigma^{\prime}}\left(\mathbf{k}\right)=-\sum_{\mathbf{k^{\prime}}}V_{\mathbf{k,k^{\prime}}}\bra{G}f_{\mathbf{k}^{\prime}\sigma}^{\dagger}f_{-\mathbf{k}^{\prime}\sigma^{\prime}}^{\dagger}\ket{G}
    \numberthis
\end{align*}
where $\ket{G}$ is the ground state and the expression comes from applying mean field theory to the quartic fermionic interaction term $\sum_{\boldsymbol{k},\boldsymbol{k}^{\prime}}V_{\boldsymbol{k},\boldsymbol{k}^{\prime}}f_{\boldsymbol{k}\sigma}^{\dagger}f_{-\mathbf{k}\sigma^{\prime}}^{\dagger}f_{\boldsymbol{k}^{\prime}\sigma}f_{-\mathbf{k}^{\prime}\sigma^{\prime}}$ where $\sigma$ indicates up or down spin, $f^\dagger_{{\bf{k}}\sigma}$'s are fermionic creation operators for momentum $\bf{k}$ and spin $\sigma$ and $V_{\bf{k},\bf{k^\prime}}$ is the Fourier coefficient of the interaction term. Thus, the superconducting order parameter $\triangle_{\sigma\sigma^{\prime}}\left(\mathbf{k}\right)$ is a wave-function of a Cooper pair formed by two quasi-particles whose
momenta and spins are $\left(\mathbf{k},\sigma\right)$ and $\left(-\mathbf{k},\sigma^{\prime}\right)$. In the s-wave superconductors, we have $\triangle_{\up\up}\left(\mathbf{k}\right)=\triangle_{\down\down}\left(\mathbf{k}\right)=0$
and 
\[
\triangle_{\up\down}\left(\mathbf{k}\right)=-\triangle_{\down\up}\left(\mathbf{k}\right)=-\sum_{\mathbf{k^{\prime}}}V_{\mathbf{k,k^{\prime}}}\bra{G}f_{\mathbf{k}^{\prime}\up}^{\dagger}f_{-\mathbf{k}^{\prime}\down}^{\dagger}\ket{G}
\numberthis
\]
In p-wave superconductors, the spatial part of the pair wave function is anti-symmetric. Thus, the spins pair up as triplets. Magnitude of total spin of a Cooper pair is $S=1$ in spin-triplet
pairing in contrast to $S=0$ in spin-singlet pairing. The spin-triplet pairing order parameter matrix has $\triangle_{\up\down}(\textbf{k})=\triangle_{\down\up}(\textbf{k})$. 
Because the order parameter and the Cooper pair wave function have the
same symmetries, the state vector $\ket{\psi}$ of a triplet superconductor
is written as 
\begin{align*}
 & \ket{\psi}\\
  = & \triangle_{\up\up}\left(\mathbf{k}\right)\ket{\up\up}+\triangle_{\down\down}\left(\mathbf{k}\right)\ket{\down\down}+\triangle_{\down\up}\left(\mathbf{k}\right)\left(\ket{\up\down}+\ket{\down\up}\right)\\
  = & \triangle_{\up\up}\left(\mathbf{k}\right)\ket{S_z=1}+\triangle_{\down\down}\left(\mathbf{k}\right)\ket{S_z=-1}+\triangle_{\down\up}\left(\mathbf{k}\right)\ket{S_z=0}
  \numberthis
\end{align*}
where the eigenstates are labeled by $S_z=-1,0,1$ where $S_z$ is the eigenvalue of the $z$-component of total spin operator of the Cooper pair. The state vector can be written in a new basis as 
\begin{align*}
    & \ket{\psi}\\
  = & \triangle_{\up\up}\left(\mathbf{k}\right)\ket{\up\up}+\triangle_{\down\down}\left(\mathbf{k}\right)\ket{\down\down}+\triangle_{\down\up}\left(\mathbf{k}\right)\left(\ket{\up\down}+\ket{\down\up}\right)\\
  = & \frac{1}{\sqrt{2}}d_{x}\left(\mathbf{k}\right)\left(-\ket{\up\up}+\ket{\down\down}\right)+\frac{i}{\sqrt{2}}d_{y}\left(\mathbf{k}\right)\left(\ket{\up\up}+\ket{\down\down}\right)+\frac{1}{\sqrt{2}}d_{z}\left(\mathbf{k}\right)\left(\ket{\up\down}+\ket{\down\up}\right)\\
 = & d_{x}\left(\mathbf{k}\right)\ket{S_{x}=0}+id_{y}\left(\mathbf{k}\right)\ket{S_{y}=0}+d_{z}\left(\mathbf{k}\right)\ket{S_{z}=0}.
 \numberthis
\end{align*}
Here, the state vector is expressed in terms of a unit vector in spin space, $\hat{d}$ on which the projection of Cooper pair spin is zero. We can write the order parameter in terms of components of $\hat{d}$ as 
\begin{align*}
\triangle\left(\mathbf{k}\right) = \left(\begin{array}{cc}
\triangle_{\up\up}\left(\mathbf{k}\right) & \triangle_{\down\up}\left(\mathbf{k}\right)\\
\triangle_{\down\up}\left(\mathbf{k}\right) & \triangle_{\down\down}\left(\mathbf{k}\right)
\end{array}\right)
=\frac{1}{\sqrt{2}}\left(\begin{array}{cc}
-d_{x}({\bf{k}})+i d_{y}({\bf{k}}) & d_{z}(\bf{k})\\
d_{z}(\bf{k}) & d_{x}({\bf{k}})+id_{y}({\bf{k}})
\end{array}\right).
\numberthis
\end{align*}
For a $p_x+ip_y$ superconductor, we consider that all the components of order parameter matrix have the same $\bf{k}$-dependence i.e. $d_i({\bf{k}})= d_i(k_x+ik_y)$ such that
\begin{align*}
\triangle_{p+ip}\left(\mathbf{k}\right)=\frac{1}{\sqrt{2}}\left(\begin{array}{cc}
-d_{x}+i d_{y} & d_{z}\\
d_{z} & d_{x}+id_{y}
\end{array}\right)(k_x+ik_y),
\numberthis
\label{order_pm}
\end{align*} where the entries in the matrix are constants. The Hamiltonian of the spin-triplet superconductor with the above order parameter matrix can be written as 
\begin{align*}
\hat{H}=\sum_{\boldsymbol{k}}\left(\begin{array}{cc}
f_{\boldsymbol{k}\up}^{\dagger} & f_{\boldsymbol{k}\down}^{\dagger}\end{array}\right)\left(\begin{array}{cc}
\xi_{\boldsymbol{k}\sigma} & 0\\
0 & \xi_{\boldsymbol{k}\sigma}
\end{array}\right)\left(\begin{array}{c}
f_{\boldsymbol{k}\up}\\
f_{\boldsymbol{k}\down}
\end{array}\right)+\sum_{\boldsymbol{k}}\left[\left(\begin{array}{cc}
f_{\boldsymbol{k}\up}^{\dagger} & f_{\boldsymbol{k}\down}^{\dagger}\end{array}\right)\triangle_{p+ip}\left(\mathbf{k}\right)\left(\begin{array}{c}
f_{-\boldsymbol{k}\up}^{\dagger}\\
f_{-\boldsymbol{k}\down}^{\dagger}
\end{array}\right)+h.c.\right]
\numberthis
\label{Ham_spin_Trip}
\end{align*}
where $\xi_{\boldsymbol{k}\sigma}$ is the single-particle kinetic energy and $h.c.$ denotes the Hermitian conjugate of the term in the bracket. Conventionally, a unitary state ($\vec{d}({\bf{k}})\times \vec{d}({\bf{k}})^\star=0$) is considered to describe a $p+ip$ superconductor. Here, we take a more general form \cref{order_pm} which can be diagonalized such that the corresponding transformation matrix is momentum independent. Hence, with the change of choice of the spin-quantization axis, the Hamiltonian \cref{Ham_spin_Trip} can be diagonalized and written as  

\begin{align*}
    H=\sum_{\boldsymbol{k}\sigma}\xi_{\boldsymbol{k}\sigma}f_{\boldsymbol{k}\sigma}^{\dagger}f_{\boldsymbol{k}\sigma}+\frac{1}{2}\left(\tilde{\triangle}_{\sigma\sigma}(\boldsymbol{k})^{\star}f_{-\boldsymbol{k}\sigma}f_{\boldsymbol{k}\sigma}+\tilde{\triangle}_{\sigma\sigma}(\boldsymbol{k})f_{\boldsymbol{k}\sigma}^{\dagger}f_{-\boldsymbol{k}\sigma}^{\dagger}\right)
    \numberthis
    \label{diag_ham}
\end{align*}
where $\xi_{\boldsymbol{k}\sigma}$ is the single-particle kinetic energy and the $\tilde{\triangle}_{\sigma\sigma}({\bf{k}})$ are the components of the order parameter matrix in the new spin coordinate system or choice of spin-quantization axis. Note that the transformation to diagonalize the Hamiltonian matrix is not momentum dependent because the momentum dependence from the off-diagonal components is separated out as $k_x+ik_y$. Ground state of this Hamiltonian is expressed in equation \cref{gs_pwave1}. 

\section{Spin-triplet Josephson junction and half-fluxon solution}
\label{half_fluxon_solution}
The Hamiltonian density for the spin-triplet Josephson junction, with the coefficients as assumed in the main text, is given by
\begin{align*}
{\cal{H}}= &\sum_{\sigma=\up,\down}\frac{1}{2}c_{n}n_{\sigma}^{2}+ \frac{1}{2}c_{\Theta}\left(\pd_{x}\Theta^{\sigma}\right)^{2} + J\left(1-\cos\Theta^{\sigma}\right)-\frac{I_b}{2}\Theta^\sigma,
\numberthis\label{ham_density}
\end{align*}
where $\Theta^\sigma$ is the order parameter phase difference across the junction for spin-component $\sigma$ and $n_\sigma$ is the number density for spin $\sigma$. Here, $c_n$ is the coefficient of the capacitive terms, $c_\Theta$ is the coefficient of the magnetic terms, $J$ is the Josephson energy scale and $I_b$ is the bias current. The corresponding Lagrangian density is given by
\begin{align*}
    {\cal{L}}= &\sum_{\sigma=\up,\down} \frac{1}{2}c_{n}\left(\pd_{t}\Theta^{\sigma}\right)^{2} - \frac{1}{2}c_{\Theta}\left(\pd_{x}\Theta^{\sigma}\right)^{2}- J\left(1-\cos\Theta^{\sigma}\right)+\frac{I_b}{2}\Theta^\sigma.
\numberthis\label{lag_density1}
\end{align*}
The momentum conjugate to $\Theta^\sigma$ is the number density of spin $\sigma$, $n_{\sigma}= \frac{\pd {\cal{L}}}{\pd({\pd_t\Theta^\sigma})}=c_n \pd_t \Theta^\sigma$ such that the total number density $n=n_\up+n_\down= c_n\pd_t(\Theta^\up+\Theta^\down)$. It is the phase conjugate to the total particle number density that couples to the magnetic vector potential. In order to find this phase, it is convenient to express $\Theta^\sigma$ as $\Theta^\up=\phi-\theta$ and $\Theta^\down=\phi+\theta$ such that $\phi$ can be factored out as a common phase for the diagonal order parameter matrix as used in \cref{diag_ham}. The Lagrangian density can then be expressed as 
\begin{align*}
  {\cal{L}}=&c_{n}\left(\pd_{t}\phi\right)^{2}+ c_{n}\left(\pd_{t}\theta\right)^{2}- c_{\Theta}\left(\pd_{x}\phi\right)^{2}- c_{\Theta}\left(\pd_{x}\theta\right)^{2}
-2J\left(1-\cos\phi\cos\theta\right)+I_b\phi.
\numberthis\label{lag_density2}
\end{align*} 
The momentum conjugate to $\phi$ is given by $\frac{\pd {\cal{L}}}{\pd({\pd_t\phi})}=2c_n\pd_t\phi$ which is the same as the total particle density $n$. The half-fluxon solution mentioned in \cref{hf_soln} is given by $\phi_{\text{HF}}= \frac{1}{2}(\Theta^\up+\Theta^\down)= 2 \arctan(e^{\pm\frac{\bar{x}-u\bar{t}}{\sqrt{1-u^2}}}) $. Following the relation~\cite{JJ_physics_book,HF_chiral} of the magnetic field to the phase that couples to the vector potential, $\phi$, we get the magnetic flux $\Phi_{\text{HF}}$ due to the half-fluxon as
\[
\Phi_{\text{HF}}  =\frac{\hbar}{2e} \int_{-\infty}^{\infty} (\pd_x\phi_{\text{HF}}) dx\\
 = \frac{h}{4e}
\numberthis
\]
which is one half of the flux quantum $\frac{h}{2e}$. 

\section{Tunnel creation of pair of half fluxon and anti-half-fluxon}
\label{tunnel_creation_app}
In this section, we study tunnel creation of an HF/AHF pair in a spin-triplet Josephson junction. As mentioned in the main text, a defect made of a localized dipole current facilitates creation of an HF-AHF pair such that the HF compensates the magnetic flux of the pinned defect while the AHF is free to move along the length of the junction. Such a pair creation can happen via under-barrier tunneling, starting from a vacuum configuration. In the presence of the defect, the vacuum configuration is not a homogeneous solution $\phi=0$ but one of the static inhomogeneous vacuum configurations as discussed below in Sec.~\ref{inhom_vac}. Starting from this inhomogeneous vacuum configuration, we consider an instanton solution, that under imaginary time evolution ends up on the mass shell as a pair configuration of localized HF (LHF) and free AHF (FAHF). In sec.~\ref{pair_rates}, we find a critical value of separation between the HF and the AHF, $z_\text{cr}$ at which the pair configurations comes on shell and use that to calculate the action corresponding to the under-barrier trajectory and the pair-creation rate. We also calculate the pair-creation rate for a configuration of localized fluxon (LF) and free anti-fluxon (FAF) and compare it with the LHF/FAHF pair-creation rate. We follow the references~\cite{ROMP_solitons, pair_creation, setup} for these calculations. From now, without loss of generality, we set the coefficients $J$, $c_\Theta$ and $c_n$ to be 1. 

\subsection{Static vacuum configurations}
\label{inhom_vac}
The static equations of motion associated with the Hamiltonian density \cref{ham_density} for $I_b=0$ are given by 
\begin{align*}
-\Theta^\up_{xx}+\sin\Theta^\up&=-\varepsilon\delta^{\prime}\left(x\right)\\
-\Theta^\down_{xx}+\sin\Theta^\down &=-\varepsilon\delta^{\prime}\left(x\right),
\numberthis\label{eqn_defect}
\end{align*}
where the time derivative term has been set to 0 to look for static solutions. Depending on the value of the defect strength the vacuum solution $\Theta^{\up(\down)}$ can be $\pm 4 \sgn\left(x\right)\arctan\left(\exp\left(\xi^{\up(\down)}-\abs x\right)\right)$ or $4\arctan\left(\exp\left(x-\xi^{\up(\down)}\sgn\left(x\right)\right)\right)$ where $\sgn\left(x\right)$ is the \textit{sign} function and the parameters $\xi^{\up(\down)}$ are fixed by the boundary conditions. We consider these solutions one by one. 

\subsubsection{Quadrupole solution}
The quadrupole solution is given by 
\begin{align*}
\Theta^\up_{\text{qd}}\left(x\right)=&\pm4\sgn\left(x\right)\arctan\left(\exp\left(\xi^{\up \pm}-\abs x\right)\right)\\
\Theta^\down_{\text{qd}}\left(x\right)=&\pm4\sgn\left(x\right)\arctan\left(\exp\left(\xi^{\down\pm}-\abs x\right)\right),
\numberthis\label{eqnsolqp}
\end{align*}
where the $\pm$ signs indicate the two allowed solutions with $+$ and $-$ signs in front. The boundary conditions~\cref{pert_bc_eqn} can be found by integrating the equations of motion \cref{eqn_defect} around $x=0$. By plugging the solutions $\Theta^\up_\text{qd}$ and $\Theta^\down_\text{qd}$ into the boundary conditions $\Theta^{\up(\down)}\left(x=+0\right)-\Theta^{\up(\down)}\left(x=-0\right)=-\varepsilon$, we get 
\begin{align*}
e^{\xi^\pm}=&\pm \tan\left(\frac{\varepsilon}{8}\right).
\numberthis\label{eqnxipm}
\end{align*}
where $\xi^\pm=\xi^{\up\pm}=\xi^{\down\pm}$ and \cref{eqnxipm} shows that $\xi^+$ and $\xi^-$ differ only by a sign. Hence, the solutions \cref{eqnsolqp} can be expressed in terms of the defect strength $\varepsilon$ as 
\begin{align*}
\Theta_\text{qd}^{\up(\down)}\left(x\right)=&\pm 4\sgn\left(x\right)\arctan\left(\pm\tan\left(\frac{\varepsilon}{8}\right)e^{-\abs x}\right)
=4\sgn\left(x\right)\arctan\left(\tan\left(\frac{\varepsilon}{8}\right)e^{-\abs x}\right)
\numberthis\label{eqn}
\end{align*}
The energy of the quadrupole configuration can be calculated as 
\begin{align*}
& E_\text{qd}\\
= & \frac{1}{2}\int_{x}\left(\pd_{x}\Theta^\up_{\text{qd}}\left(x\right)\right)^{2}+\frac{1}{2}\int_x\left(\pd_{x}\Theta^\down_{\text{qd}}\left(x\right)\right)^{2}+\int_{x}\left(1-\frac{1}{2}\cos\Theta^\up_{\text{qd}}-\frac{1}{2}\cos\Theta^\down_{\text{qd}}\right)-\frac{1}{2}\int_{x}\varepsilon\delta^{\prime}\left(x\right)(\Theta^\up_{\text{qd}}\left(x\right)+\Theta^\down_{\text{qd}}\left(x\right))\\
=& \int_{x}\left(\pd_{x}\Theta^\up_{\text{qd}}\left(x\right)\right)^{2}+\int_{x}\left(1-\cos\Theta^\up_{\text{qd}}\right)-\int_{x}\varepsilon\delta^{\prime}\left(x\right)\Theta^\up_{\text{qd}}\left(x\right)\\
=& 2 \left(\varepsilon \sin \left(\frac{\varepsilon}{4}\right)-6 \cos\left(\frac{\varepsilon}{4}\right)+6\right)
\numberthis\label{vac_energy}
\end{align*}
which goes to $0$ as $\varepsilon\rightarrow 0$ leads to the homogeneous solution $\Theta^{\up(\down)}=0$.

\subsubsection{Localized half fluxon solution}
A localized half fluxon (LHF) solution localized at the defect in the presence of the defect potential $-\varepsilon \delta^\prime(x)$, can be written~\cite{pair_creation} as 
\begin{align*}
\Theta^{\up}_{\text{LHF}}\left(x\right)=&4\arctan\left(\exp\left(x-\xi_{\text{LHF}}^{\up}\sgn\left(x\right)\right)\right)\\
\Theta^{\down}_{\text{LHF}}\left(x\right)=&\pm4\sgn\left(x\right)\arctan\left(\exp\left(\xi^\down_{\text{LHF}}-\abs x\right)\right)
\numberthis\label{HFsoln}
\end{align*}
Note that $\Theta^\up_{\text{LHF}}$ goes from $0$ at $x\rightarrow-\infty$ to $2\pi$ at $x\rightarrow\infty$ while $\Theta^\down_{\text{LHF}}$ doesn't get a net phase jump. By plugging this solution into the boundary conditions $\Theta^{\up(\down)}\left(x=+0\right)-\Theta^{\up(\down)}\left(x=-0\right)=-\varepsilon$, we get 
\begin{align*}
e^{\xi_{\text{LHF}}^\up}=&\frac{1+\tan\left(\frac{\varepsilon}{8}\right)}{1-\tan\left(\frac{\varepsilon}{8}\right)}\\
e^{\xi_{\text{LHF}}^\down}=&\pm \tan{\frac{\varepsilon}{8}}.
\numberthis\label{eqnxiLHF}
\end{align*}
Hence, the energy of the localized half-fluxon configuration can be expressed in terms of $\Theta^\up_{\text{LHF}}$ and $\Theta^\down_{\text{LHF}}$ and calculated in terms of the defect strength $\varepsilon$ as 
\begin{align*}
& E_{\text{LHF}}\\
= & H\left(\pd_{t}\Theta^\up_{\text{LHF}},\pd_{x}\Theta^\up_{\text{LHF}},\pd_{t}\Theta^\down_{\text{LHF}},\pd_x\Theta^\down_{\text{LHF}}\right)\\
= & \int_{x}\frac{1}{2}\left(\pd_{x}\Theta^\up_{\text{LHF}}\right)^{2}+\frac{1}{2}\left(\pd_{x}\Theta^\down_{\text{LHF}}\right)^{2}+\int_{x}\Big(\left(1-\frac{1}{2}\cos\Theta^\up_{\text{LHF}}-\frac{1}{2}\cos\Theta^\down_{\text{LHF}}\right)-\frac{1}{2}\varepsilon\delta^{\prime}\left(x\right)(\Theta^\up_{\text{LHF}}\left(x\right)+\Theta^\down_{\text{LHF}}\left(x\right))\Big)\\
= & (\varepsilon +6) \sin \left(\frac{\varepsilon
   }{4}\right)+(\varepsilon -6) \cos \left(\frac{\varepsilon
   }{4}\right)+12
\numberthis\label{eqnHam_HF}
\end{align*}

\subsubsection{Localized fluxon solution}
A localized fluxon(LF) solution for $\Theta^\up$ and $\Theta^\down$ in the presence of the defect potential $-\varepsilon \delta^\prime(x)$ can be written~\cite{pair_creation} as 
\begin{align*}
\Theta^{\up}_{\text{LF}}\left(x\right)=&4\arctan\left(\exp\left(x-\xi_{\text{LF}}^{\up}\sgn\left(x\right)\right)\right)\\
\Theta^{\down}_{\text{LF}}\left(x\right)=&4\arctan\left(\exp\left(x-\xi_{\text{LF}}^{\up}\sgn\left(x\right)\right)\right)
\numberthis\label{eqn}
\end{align*}
Note that both $\Theta^\up_{\text{LF}}$ and $\Theta^\down_{\text{LF}}$ go from $0$ at $x\rightarrow-\infty$ to $2\pi$ at $x\rightarrow\infty$. By plugging this solution into the boundary conditions $\Theta^{\up(\down)}\left(x=+0\right)-\Theta^{\up(\down)}\left(x=-0\right)=-\varepsilon$, we get 
\begin{align*}
e^{\xi_{\text{LF}}^{\up(\down)}}=&\frac{1+\tan\left(\frac{\varepsilon}{8}\right)}{1-\tan\left(\frac{\varepsilon}{8}\right)}
\numberthis\label{eqnxiLHF}
\end{align*}
Hence, the energy of the localized fluxon configuration can be expressed in terms of $\Theta^\up_{\text{LF}}$ and $\Theta^\down_{\text{LF}}$ as 
\begin{align*}
& E_{\text{LF}}\\
= & H\left(\pd_{t}\Theta^\up_{\text{LF}},\pd_{x}\Theta^\up_{\text{LF}},\pd_{t}\Theta^\down_{\text{LF}},\pd_x\Theta^\down_{\text{LF}}\right)\\
= & \int_{x}\frac{1}{2}\left(\pd_{x}\Theta^\up_{\text{LF}}\right)^{2}+\frac{1}{2}\left(\pd_{x}\Theta^\down_{\text{LF}}\right)^{2}+\int_{x}\Big(\left(1-\frac{1}{2}\cos\Theta^\up_{\text{LF}}-\frac{1}{2}\cos\Theta^\down_{\text{LF}}\right)-\frac{1}{2}\varepsilon\delta^{\prime}\left(x\right)(\Theta^\up_{\text{LF}}\left(x\right)+\Theta^\down_{\text{LF}}\left(x\right))\Big)\\
= & 2\varepsilon\cos\frac{\varepsilon}{4}+ 12\sin{\frac{\varepsilon}{4}}+12
\numberthis\label{eqnHam_HF}
\end{align*}

\begin{figure}
\centering
\includegraphics[scale=1.3]{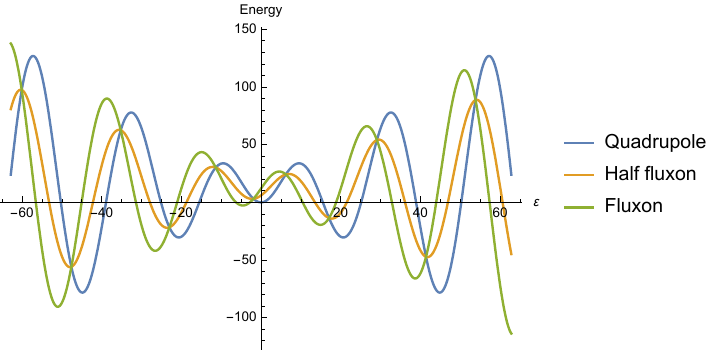}
\vspace{3mm}
\caption{Energy of possible vacuum configurations as a function of defect strength $\varepsilon$. The energy for the half-fluxon solution stays in between the quadrupole and the fluxon solution which are the two stable vacuum configurations. A defect strength can be chosen such that one of these configurations is the vacuum configuration.}
\label{vacuum}
\vspace{-5mm}
\end{figure}

It turns out that the stable vacuum configuration can be either the quadrupole solution or the localized fluxon solution, as can be seen from Fig.~\ref{vacuum}. We assume that a value of the defect strength is chosen such that the vacuum configuration is the quadrupole solution. 

\subsection{Localized half fluxon/free anti-half-fluxon (LHF/FAHF) pair configuration}
\label{pair_creation}
Consider an HF-AHF pair configuration such that the HF is localized around the defect at $x=0$ and the AHF is free and far away from the defect. Due to the separation of FAHF from the defect/LHF, we can take the defect strength near FAHF to be $0$ and assume that there is no interaction between the FAHF and the defect or LHF. Hence, the FAHF solution can be written as 
\begin{align*}
\Theta^{\up}_{\text{FAHF}}\left(x,t\right)=&4\arctan\left(\exp\left(\beta^{-1}\left(-x+z(t)\right)\right)\right)\\
\Theta^{\down}_{\text{FAHF}}\left(x,t\right)=&0
\numberthis\label{eqn}
\end{align*}
where $\beta$ is the Lorentzian factor, which is a function of velocity of FAHF, $\frac{dz}{dt}$ and is given by $\beta=\sqrt{1-(\frac{dz}{dt})^{2}}$.
From the solutions for LHF and FAHF, we can write the LHF/FAHF under-barrier pair configuration as
\begin{align*}
& \Theta^{\up}_{\text{LHF,FAHF}}\left(x,\tau=it\right)=-2\pi+4\arctan\left(\exp\left(x-\xi^\up_{\text{LHF}}\sgn\left(x\right)\right)\right)+4\arctan\left(\exp\left(\frac{-x+z}{\beta(\dot{z})}\right)\right)\\
& \Theta^{\down}_{\text{LHF,FAHF}}\left(x,\tau=it\right)=\pm4\sgn\left(x\right)\arctan\left(\exp\left(\xi^\down_{\text{LHF}}-\abs x\right)\right)
\numberthis\label{eqn}
\end{align*}
where $\xi^{\up(\down)}_{\text{LHF}}$ obey \cref{eqnxiLHF} and $\beta\left(\dot{z}\right)=\sqrt{1+\dot{z}^{2}}$ with $\dot{z}=\frac{dz}{d\tau}$ where $\tau=it$ is imaginary time. Substituting this in the Hamiltonian and taking large LHF-FAHF separation, we get the energy gap of the pair configuration w.r.t. the quadrupole vacuum energy \cref{eqnHam_HF} as
\begin{align*}
& E^{\text{qd}}_{\text{LHF,FAHF}}\\
= & E_{\text{LHF,FAHF}}-E_{\text{qd}}\\
= & \frac{1}{2}\int_{x}\Big[\left(\pd_{x}\Theta^\up_{\text{LHF}}\right)^{2}+\left(\pd_{t}\Theta^\up_{\text{LHF}}\right)^{2}+\left(\pd_{x}\Theta^\down_{\text{LHF}}\right)^{2}+\left(\pd_{t}\Theta^\down_{\text{LHF}}\right)^{2}\\
& +\left(\pd_{x}\Theta^\up_{\text{FAHF}}\right)^{2}+\left(\pd_{t}\Theta^\up_{\text{FAHF}}\right)^{2}+\left(\pd_{x}\Theta^\down_{\text{FAHF}}\right)^{2}+\left(\pd_{t}\Theta^\down_{\text{FAHF}}\right)^{2}\Big]+\int_{x}\Big(\left(1-\frac{1}{2}\cos\Theta^\up_{\text{LHF,FAHF}}-\frac{1}{2}\cos\Theta^\down_{\text{LHF,FAHF}}\right)\\
& -\frac{1}{2}(I_b+\varepsilon\delta^{\prime}\left(x\right))(\Theta^\up_{\text{LHF,FAHF}}\left(x\right)+\Theta^\down_{\text{LHF,FAHF}}\left(x\right))\Big)-E_{\text{qd}}\\
= & (6-\varepsilon)\sin\frac{\varepsilon}{4}+(6+\varepsilon)\cos\frac{\varepsilon}{4}+2\left(\frac{4-\beta(\dot{z})^2}{\beta(\dot{z})}\right)-I_b \pi z .
\numberthis\label{eqnHam_HF}
\end{align*}
where we used \cref{eqnxiLHF} in the calculation and took $\beta(\dot{z})$ or the velocity of FAHF to be a constant. For the bias current contribution, we have approximated the pair profile by step functions at $x=0,z$ such that $\Theta^{\up}_{\text{LHF,FAHF}}$ is equal to $2\pi$ only in the region $0<x<z$ and $0$ elsewhere.

\subsection{Localized fluxon/free anti-fluxon (LF/FAF) pair configuration}
We can write the LF/FAF pair configuration as 
\begin{align*}
\Theta^{\up(\down)}_{\text{LF,FAF}}\left(x,\tau=it\right)=-2\pi+4\arctan\left(\exp\left(x-\xi_{\text{LF}}\sgn\left(x\right)\right)\right)+4\arctan\left(\exp\left(\frac{-x+z}{\beta(\dot{z})}\right)\right)
\numberthis\label{eqn2LHF}
\end{align*}
where both the phases $\Theta^\up=\Theta^\down$ jump by $2\pi$. By plugging this solution into the boundary conditions $\Theta^{\up(\down)}\left(x=+0\right)-\Theta^{\up(\down)}\left(x=-0\right)=-\varepsilon$, we get 
\begin{align*}
e^{\xi_{\text{LF}}^{\up(\down)}}=&\frac{1+\tan\left(\frac{\varepsilon}{8}\right)}{1-\tan\left(\frac{\varepsilon}{8}\right)}
\numberthis\label{eqnxi2LHF}
\end{align*}
Substituting \cref{eqn2LHF} in the Hamiltonian and taking large LF/FAF separation, we get the energy gap of the pair configuration LF/FAF w.r.t. the quadrupole vaccum as 
\begin{align*}
&E^{\text{qd}}_{\text{LF,FAF}}\\
= &  E_{\text{LF,FAF}}-E_{\text{qd}}\\
=& \int_{x}\Big(\left(\pd_{x}\Theta^\up_{\text{LF}}\right)^{2}+\left(\pd_{x}\Theta^\up_{\text{FAF}}\right)^{2}+\left(\pd_{t}\Theta^\up_{\text{FAF}}\right)^{2}+\int_{x}\left(1-\frac{1}{2}\cos\Theta^\up_{\text{LF,FAF}}-\frac{1}{2}\cos\Theta^\down_{\text{LF,FAF}}\right)-I_b\Theta^\up_{\text{LF,FAF}}-\varepsilon\delta^{\prime}\left(x\right)\Theta^\up_{\text{LF}}\Big)-E_{\text{qd}}\\
=& 2\varepsilon\cos\frac{\varepsilon}{4}+ 12\sin{\frac{\varepsilon}{4}}+12- \left((\varepsilon +6) \sin \left(\frac{\varepsilon
   }{4}\right)+(\varepsilon -6) \cos \left(\frac{\varepsilon
   }{4}\right)+12\right)+4\frac{4-\beta^2}{\beta}-2I_b\pi z\\
=& 2(6-\varepsilon) \sin \left(\frac{\varepsilon
   }{4}\right)+2(6+\varepsilon) \cos \left(\frac{\varepsilon
   }{4}\right)+4\frac{4-\beta^2}{\beta}-2I_b\pi z\\
= & 2\, E^v_{\text{LHF,FAHF}}.
\numberthis\label{eqn}
\end{align*}
where we used \cref{eqnxi2LHF} in the calculation.

\subsection{Pair creation rates}\label{pair_rates}
Now we calculate the pair creation rates for LHF/FAHF and LF/FAF pair configurations. The Lagrangian density ${\cal L}$ is given by
\begin{align*}
{\cal L}&=\sum_\sigma-\frac{1}{2}\left(\pd_{\tau}\Theta^\sigma\right)^{2}-\frac{1}{2}\left(\pd_{x}\Theta^\sigma\right)^{2}-\left(1-\frac{1}{2}\cos\Theta^\sigma\right)+\frac{I_b+\varepsilon\delta^\prime\left(x\right)}{2}\Theta^\sigma\\
&= -{\cal{H}}-\sum_\sigma\left(\pd_{\tau}\Theta^\sigma\right)^{2}
\numberthis\label{lag_dens}
\end{align*}
where ${\cal H}$ is the Hamiltonian density,  
\begin{align*}
{\cal H}&=-\frac{1}{2}\left(\pd_{\tau}\Theta^\sigma\right)^{2}+\frac{1}{2}\left(\pd_{x}\Theta^\sigma\right)^{2}+\left(1-\frac{1}{2}\cos\Theta^\sigma\right)-\frac{I_b+\varepsilon\delta^\prime\left(x\right)}{2}\Theta^\sigma.
\numberthis\label{ham_dens}
\end{align*}
The effective pair configuration Lagrangian as a function of coordinates $z$ and $\dot{z}$ can be expressed in terms of the energy of pair configuration w.r.t vacuum as
\begin{align*}
L_\text{pair}(z,\dot{z})
& =-E^v_\text{pair}-\sum_\sigma\int\left(\pd_{\tau}\Theta^\sigma\right)^{2}. 
\label{eff_lag}
\numberthis
\end{align*}

\subsubsection{LHF-FAHF configuration}

Using \cref{eff_lag}, the effective Lagrangian for LHF/FAHF configuration is hence given by 
\begin{align*}
L_{\text{LHF,FAHF}}(z,\dot{z})
& =-E^v_{\text{LHF,FAHF}}-\frac{8\dot{z}^2}{\beta(\dot{z})}\\
& =-(6-\varepsilon)\sin\frac{\varepsilon}{4}-(6+\varepsilon)\cos\frac{\varepsilon}{4}-6\beta(\dot{z})+I_b \pi z
\end{align*}
For the effective Lagrangian, the momentum conjugate to the coordinate $z$ is given by $p_{\text{HF}}=\frac{\pd L_{\text{LHF,FAHF}}}{i\pd \dot{z}}=-6\frac{\dot{z}}{i\beta(\dot{z})}$. Hence, we can write the effective Hamiltonian $H_{\text{LHF,FAHF}}(p,z)$ as 
\begin{align*}
H_{\text{LHF,FAHF}}(p,z)&=ip_{\text{HF}}\dot{z}-L_{\text{LHF,FAHF}}(z,\dot{z})\\
&=(6-\varepsilon)\sin\frac{\varepsilon}{4}+(6+\varepsilon)\cos\frac{\varepsilon}{4}-6\frac{\dot{z}^2}{\beta(\dot{z})}+6\beta(\dot{z})-I_b \pi z\\
&=(6-\varepsilon)\sin\frac{\varepsilon}{4}+(6+\varepsilon)\cos\frac{\varepsilon}{4}+\frac{6}{\beta(\dot{z})}-I_b \pi z\\
&=(6-\varepsilon)\sin\frac{\varepsilon}{4}+(6+\varepsilon)\cos\frac{\varepsilon}{4}+\sqrt{p_{\text{HF}}^2+E_{\text{FAHF},(0)}^2}-I_b \pi z
\end{align*}
where $E_{\text{FAHF},(0)}=6$ is the energy of FAHF at zero velocity. For the under barrier trajectory, we can set the effective Hamiltonian $H_{\text{LHF,FAHF}}(p,z)=0$ for spontaneous pair creation from vacuum. Thus, we can write the momentum for the under-barrier trajectory as 
\[
p_{\text{HF}}=\sqrt{\left(I_b\pi z-(6-\varepsilon)\sin\frac{\varepsilon}{4}-(6+\varepsilon)\cos\frac{\varepsilon}{4}\right)^2-36}.
\]
Action corresponding to the under-barrier trajectory for $H(p_{\text{HF}},z)=0$ is given by $S\left(\triangle E\right)=\int_{0}^{z_\text{cr}}dt (ip_{\text{HF}}\dot{z}) =-i\int_{0}^{z_\text{cr}}d\tau (ip_{\text{HF}}\frac{dz}{d\tau}) 
=\int_{0}^{z_\text{cr}}dz p_{\text{HF}}$ where $z_\text{cr}$ is the HF-AHF separation at which the pair configuration becomes on-shell. $z_\text{cr}$ is defined via $H^{eff}_{\text{LHF,FAHF}}\left(p_{\text{HF}}=0,z_\text{cr}\right)=0$ and using this, we get
\begin{align*}
z_\text{cr} &=\frac{(6-\varepsilon)\sin\frac{\varepsilon}{4}+(6+\varepsilon)\cos\frac{\varepsilon}{4}+6}{I_b\pi},
\numberthis\label{eqn}
\end{align*}
which makes sense because $z_\text{cr}I_b\pi$ is the energy gain that compensates the pair energy which is $12$ for $\varepsilon,I_b=0$. Thus, we get the effective Euclidean action as 
\begin{align*}
& S^{Euc}_{\text{LHF,FAHF}}\\
= & i\int_{0}^{z_\text{cr}}dz p_{\text{HF}}\\
= & \int_{0}^{z_\text{cr}}dz \sqrt{\left(I_b\pi z-(6-\varepsilon)\sin\frac{\varepsilon}{4}-(6+\varepsilon)\cos\frac{\varepsilon}{4}\right)^2-36}.
\end{align*}

\subsubsection{LF/FAF configuration} \label{euclidean}
Using \cref{lag_dens}, the effective Lagrangian for LF/FAF configuration is given by 
\begin{align*}
L_{\text{LF,FAF}}(z,\dot{z})
& =- E^v_{\text{LF,FAF}}-\frac{16\dot{z}^2}{\beta(\dot{z})}\\
& =- 2\left(E^v_{\text{LHF,FAHF}}-\frac{8\dot{z}^2}{\beta(\dot{z})}\right)\\
& =2 \, L_{\text{LHF,FAHF}}(z,\dot{z})
   \numberthis
   \label{lag_Rel}
\end{align*}
The momentum conjugate to coordinate $z$ is given by $p_{\text{F}}=\frac{\pd L_{\text{LF,FAF}}}{\pd \dot{z}}=-12 \frac{\dot{z}}{i\beta(\dot{z})}=2p_{\text{HF}}$. Hence, we can write the effective Hamiltonian as 
\begin{align*}
H_{\text{LF,FAF}}(p,z)&=ip_{\text{F}}\dot{z}-L_{\text{LF,FAF}}(z,\dot{z})\\
&=2\left((6-\varepsilon) \sin \left(\frac{\varepsilon
   }{4}\right)+(6+\varepsilon) \cos\left(\frac{\varepsilon
   }{4}\right)+\sqrt{p^2+6^2}-I_b\pi z\right) \\ 
&= 2\, H_{\text{LHF,FAHF}}(p_{\text{HF}},z)
\numberthis
\label{ham_rel}
\end{align*}
Following the calculation in the previous subsection for the LHF/FAHF configuration and using the above results \cref{lag_Rel} and \cref{ham_rel}, we get the critical separation for LF/FAF pair creation to be the same as that for LHF/FAHF pair creation. Thus, we get the relation between the effective Euclidean actions as
\begin{align*}
S^{Euc}_{\text{LF,FAF}}=  2 \, S^{Euc}_{\text{LHF,FAHF}} 
\numberthis
\label{relSEuc}
\end{align*}

The pair creation rate is determined by the exponentially small factor $\exp(-\frac{S^{Euc}}{\hbar})$. Hence from \cref{relSEuc}, it follows that the LF/FAF pair creation rate is exponentially suppressed compared to the rate of LHF/FAHF pair creation on top of the quadrupole vacuum configuration. 

\end{document}